\documentstyle[epsfig,a4p,12pt]{article}

\begin{document}
%
%
\newcommand{\EPnum}     {CERN-EP/98-136}
\newcommand{\PRnum}     {OPAL Paper PR256}
\newcommand{\Date}      {18th August 1998}
\newcommand{\Author}    {S.~Asai, A.~F\"{u}rtjes, M.~Gruw\'e, J.~Kanzaki,
S.~Komamiya, D.~Liu, G.~Mikenberg, K.~Nagai, 
H.~Neal, and R.~Van Kooten}
\newcommand{\MailAddr}  {Magali.Gruwe@cern.ch and Sachio.Komamiya@cern.ch}
\newcommand{\EdBoard}   {R.~McPherson, C.~Sbarra, P.~Krieger,
and R.~Teuscher} 
\newcommand{\DraftVer}  {Version 3.1}
\newcommand{\DraftDate} {17th August 1998}
\newcommand{\TimeLimit} {17th August 1998, 16:00 CERN time} 
\def\toprule{\noalign{\hrule \medskip}}
\def\midrule{\noalign{\medskip\hrule }}
\def\botrule{\noalign{\medskip\hrule }}
\setlength{\parskip}{\medskipamount}
 
\newcommand{\lumi}{57}
\newcommand{\ee}{{\mathrm e}^+ {\mathrm e}^-}
\newcommand{\sq}{\tilde{\mathrm q}}
\newcommand{\sg}{\tilde{\mathrm g}}
\newcommand{\su}{\tilde{\mathrm u}}
\newcommand{\sd}{\tilde{\mathrm d}}
\newcommand{\seff}{\tilde{\mathrm f}}
\newcommand{\sele}{\tilde{\mathrm e}}
\newcommand{\sell}{\tilde{\ell}}
\newcommand{\snu}{\tilde{\nu}}
\newcommand{\smu}{\tilde{\mu}}
\newcommand{\ch}{\tilde{\chi}^\pm}
\newcommand{\chp}{\tilde{\chi}^+}
\newcommand{\chm}{\tilde{\chi}^-}
\newcommand{\chpm}{\tilde{\chi}^\pm}
\newcommand{\chmp}{\tilde{\chi}^\mp}
\newcommand{\nt}{\tilde{\chi}^0}
\newcommand{\qq}{{\mathrm q}\bar{\mathrm q}}
\newcommand{\WW}{{\mathrm W}^+ {\mathrm W}^-}
\newcommand{\gv}{\gamma^*}
\newcommand{\Wenu}{{\mathrm W e}\nu}
\newcommand{\ZZ}{{\mathrm Z} {\mathrm Z}}
\newcommand{\Zg}{{\mathrm Z} \gamma}
\newcommand{\Zgv}{{\mathrm Z} \gamma^*}
\newcommand{\Zvgv}{{\mathrm Z^*} \gamma^*}
\newcommand{\Wv}{{\mathrm W}^{*}}
\newcommand{\Wrv}{{\mathrm W}^{(*)}}
\newcommand{\Wvp}{{\mathrm W}^{*+}}
\newcommand{\Wvm}{{\mathrm W}^{*-}}
\newcommand{\Wvpm}{{\mathrm W}^{*\pm}}
\newcommand{\Wvrpm}{{\mathrm W}^{(*)\pm}}
\newcommand{\Z}{\mathrm Z}
\newcommand{\Zv}{{\mathrm Z}^{*}}
\newcommand{\Zrv}{{\mathrm Z}^{(*)}}
\newcommand{\ffbar}{{\mathrm f}\bar{\mathrm f}}
\newcommand{\nunu}{\nu \bar{\nu}}
\newcommand{\mumu}{\mu^+ \mu^-}
\newcommand{\tautau}{\tau^+ \tau^-}
\newcommand{\ellell}{\ell^+ \ell^-}
\newcommand{\nulqq}{\nu \ell {\mathrm q} \bar{\mathrm q}'}
\newcommand{\MZ}{m_{\mathrm Z}}
 
\newcommand{\gsim}{\;\raisebox{-0.9ex}
           {$\textstyle\stackrel{\textstyle >}{\sim}$}\;}
\newcommand{\lsim}{\;\raisebox{-0.9ex}{$\textstyle\stackrel{\textstyle<}
           {\sim}$}\;}

\newcommand{\degree}    {^\circ}
%
\newcommand{\Ecm}       {E_{\mathrm{cm}}}
\newcommand{\Ebeam}     {E_{\mathrm{b}}}
\newcommand{\roots}     {\sqrt{s}}
%
%
\newcommand{\thrust}    {T}
\newcommand{\nthrust}   {\hat{n}_{\mathrm{thrust}}}
\newcommand{\thethr}    {\theta_{\,\mathrm{thrust}}}
\newcommand{\phithr}    {\phi_{\mathrm{thrust}}}
\newcommand{\acosthr}   {|\cos\thethr|}
\newcommand{\thejet}    {\theta_{\,\mathrm{jet}}}
\newcommand{\acosjet}   {|\cos\thejet|}
\newcommand{\thmiss}    { \theta_{\mathrm{miss}} }
\newcommand{\cosmiss}   {| \cos \thmiss |}
%
%
\newcommand{\Evis}      {E_{\mathrm{vis}}}
\newcommand{\Rvis}      {E_{\mathrm{vis}}\,/\roots}
\newcommand{\Mvis}      {M_{\mathrm{vis}}}
\newcommand{\Rbal}      {R_{\mathrm{bal}}}
\newcommand{\pt}        {p_{\mathrm{t}}}
%
%
\newcommand{\phiacop}   {\phi_{\mathrm{acop}}}
\newcommand{\axicos}{{\mid\cos\theta_a}^{\mathrm{miss}}\mid}
%
%
\newcommand{\PhysLett}  {Phys.~Lett.}
\newcommand{\PRL} {Phys.~Rev.\ Lett.}
\newcommand{\PhysRep}   {Phys.~Rep.}
\newcommand{\PhysRev}   {Phys.~Rev.}
\newcommand{\NPhys}  {Nucl.~Phys.}
\newcommand{\NIM} {Nucl.~Instr.\ Meth.}
\newcommand{\CPC} {Comp.~Phys.\ Comm.}
\newcommand{\ZPhys}  {Z.~Phys.}
\newcommand{\IEEENS} {IEEE Trans.\ Nucl.~Sci.}
%
%
\newcommand{\OPALColl}  {OPAL Collab.}
\newcommand{\JADEColl}  {JADE Collab.}
\newcommand{\etal}      {{\it et~al.}}
\newcommand{\onecol}[2] {\multicolumn{1}{#1}{#2}}
\newcommand{\ra}        {\rightarrow}   
 

 
\begin{titlepage}
%
%
\begin{center}
    \large
     EUROPEAN LABORATORY FOR PARTICLE PHYSICS
\end{center}
\begin{flushright}
    \large
   \EPnum\\
   \Date
\end{flushright}
%
%
%
%
\begin{center}
\Large\bf\boldmath
    Search for Chargino and Neutralino Production  
  at $\sqrt{s} = 181$--$184$~GeV at LEP
\end{center}
%
%
\begin{center}
\Large
    The OPAL Collaboration \\
\end{center}
\bigskip
\bigskip
\begin{center}
\large
\end{center}
%
%
 \begin{abstract}
A search for charginos and neutralinos, 
predicted by supersymmetric theories, has been
performed using a data sample of $\lumi$~pb$^{-1}$
at centre-of-mass energies of 181--184~GeV taken
with the OPAL detector at LEP.  
No evidence for chargino or neutralino production has been found.
Upper limits on chargino and neutralino pair production 
($\chp_1 \chm_1$,$\nt_1 \nt_2$) cross-sections are obtained
as a function of the chargino mass ($m_{\chpm_1}$),
the lightest neutralino mass ($m_{\nt_1}$) 
and the second lightest neutralino mass ($m_{\nt_2}$)\@.  
For large chargino masses the limits have been improved 
with respect to the previous analyses at lower centre-of-mass energies. 
Exclusion regions at 95\% confidence level (C.L.) of parameters of the
Constrained Minimal Supersymmetric Standard Model are determined
for the case of a large universal scalar mass, $m_0$, 
implying heavy scalar fermions, 
and for the case of a small $m_0$ resulting in light
scalar fermions and giving the worst-case limits.
Within this framework and for $m_{\chpm_1} - m_{\nt_1} \geq 5$~GeV
the 95\% C.L. lower limits on $m_{\chpm_1}$ for $m_0=500$~GeV 
are 90.0 and 90.2~GeV for $\tan \beta = 1.5$ and 35 respectively.
These limits for all $m_0$ (the worst-case) are 69.1 and 65.2~GeV 
for $\tan \beta = 1.5$ and 35 respectively.
Exclusion regions are also presented for neutralino masses, including
an absolute lower limit at 95\% C.L. for the mass of the lightest neutralino 
of $30.1$~GeV for $m_0 = 500$~GeV  
($24.2$~GeV for all $m_0$),
with implications for experimental searches for the lightest neutralino as 
a dark matter candidate.
\end{abstract}
\begin{center}
\bigskip

  
\bigskip

\noindent

\begin{center}
{\bf \Large Submitted to Eur. Phys. Journal C}\\
\end{center}

\end{center}
\end{titlepage}

\begin{center}{
G.\thinspace Abbiendi$^{  2}$,
K.\thinspace Ackerstaff$^{  8}$,
G.\thinspace Alexander$^{ 23}$,
J.\thinspace Allison$^{ 16}$,
N.\thinspace Altekamp$^{  5}$,
K.J.\thinspace Anderson$^{  9}$,
S.\thinspace Anderson$^{ 12}$,
S.\thinspace Arcelli$^{ 17}$,
S.\thinspace Asai$^{ 24}$,
S.F.\thinspace Ashby$^{  1}$
% ...
}\end{center}\bigskip
\bigskip
$^{  1}$School of Physics and Astronomy, University of Birmingham,
Birmingham B15 2TT, UK
\newline
$^{  2}$Dipartimento di Fisica dell' Universit\`a di Bologna and INFN,
I-40126 Bologna, Italy
\newline
$^{  3}$Physikalisches Institut, Universit\"at Bonn,
D-53115 Bonn, Germany
\newline
$^{  4}$Department of Physics, University of California,
Riverside CA 92521, USA
\newline
$^{  5}$Cavendish Laboratory, Cambridge CB3 0HE, UK
\newline
$^{  6}$Ottawa-Carleton Institute for Physics,
Department of Physics, Carleton University,
Ottawa, Ontario K1S 5B6, Canada
\newline
$^{  7}$Centre for Research in Particle Physics,
Carleton University, Ottawa, Ontario K1S 5B6, Canada
\newline
$^{  8}$CERN, European Organisation for Particle Physics,
CH-1211 Geneva 23, Switzerland
\newline
$^{  9}$Enrico Fermi Institute and Department of Physics,
University of Chicago, Chicago IL 60637, USA
\newline
$^{ 10}$Fakult\"at f\"ur Physik, Albert Ludwigs Universit\"at,
D-79104 Freiburg, Germany
\newline
$^{ 11}$Physikalisches Institut, Universit\"at
Heidelberg, D-69120 Heidelberg, Germany
\newline
$^{ 12}$Indiana University, Department of Physics,
Swain Hall West 117, Bloomington IN 47405, USA
\newline
$^{ 13}$Queen Mary and Westfield College, University of London,
London E1 4NS, UK
\newline
$^{ 14}$Technische Hochschule Aachen, III Physikalisches Institut,
Sommerfeldstrasse 26-28, D-52056 Aachen, Germany
\newline
$^{ 15}$University College London, London WC1E 6BT, UK
\newline
$^{ 16}$Department of Physics, Schuster Laboratory, The University,
Manchester M13 9PL, UK
\newline
$^{ 17}$Department of Physics, University of Maryland,
College Park, MD 20742, USA
\newline
$^{ 18}$Laboratoire de Physique Nucl\'eaire, Universit\'e de Montr\'eal,
Montr\'eal, Quebec H3C 3J7, Canada
\newline
$^{ 19}$University of Oregon, Department of Physics, Eugene
OR 97403, USA
\newline
$^{ 20}$CLRC Rutherford Appleton Laboratory, Chilton,
Didcot, Oxfordshire OX11 0QX, UK
\newline
$^{ 22}$Department of Physics, Technion-Israel Institute of
Technology, Haifa 32000, Israel
\newline
$^{ 23}$Department of Physics and Astronomy, Tel Aviv University,
Tel Aviv 69978, Israel
\newline
$^{ 24}$International Centre for Elementary Particle Physics and
Department of Physics, University of Tokyo, Tokyo 113-0033, and
Kobe University, Kobe 657-8501, Japan
\newline
$^{ 25}$Institute of Physical and Environmental Sciences,
Brunel University, Uxbridge, Middlesex UB8 3PH, UK
\newline
$^{ 26}$Particle Physics Department, Weizmann Institute of Science,
Rehovot 76100, Israel
\newline
$^{ 27}$Universit\"at Hamburg/DESY, II Institut f\"ur Experimental
Physik, Notkestrasse 85, D-22607 Hamburg, Germany
\newline
$^{ 28}$University of Victoria, Department of Physics, P O Box 3055,
Victoria BC V8W 3P6, Canada
\newline
$^{ 29}$University of British Columbia, Department of Physics,
Vancouver BC V6T 1Z1, Canada
\newline
$^{ 30}$University of Alberta,  Department of Physics,
Edmonton AB T6G 2J1, Canada
\newline
$^{ 31}$Duke University, Dept of Physics,
Durham, NC 27708-0305, USA
\newline
$^{ 32}$Research Institute for Particle and Nuclear Physics,
H-1525 Budapest, P O  Box 49, Hungary
\newline
$^{ 33}$Institute of Nuclear Research,
H-4001 Debrecen, P O  Box 51, Hungary
\newline
$^{ 34}$Ludwigs-Maximilians-Universit\"at M\"unchen,
Sektion Physik, Am Coulombwall 1, D-85748 Garching, Germany
\newline
\bigskip\newline
$^{  a}$ and at TRIUMF, Vancouver, Canada V6T 2A3
\newline
$^{  b}$ and Royal Society University Research Fellow
\newline
$^{  c}$ and Institute of Nuclear Research, Debrecen, Hungary
\newline
$^{  d}$ and Department of Experimental Physics, Lajos Kossuth
University, Debrecen, Hungary
\newline
$^{  e}$ on leave of absence from the University of Freiburg
\newline
 
\newpage

 
\section{Introduction}
A direct search for charginos and neutralinos
predicted in SUSY theories~\cite{SUSY}
is performed using the data collected with the OPAL detector at the 
centre-of-mass energies ($\roots$) of 181--184~GeV at the 
LEP $\ee$ collider at CERN.
At these energies, 
chargino production cross-sections as large as 3.6~pb
for a mass of 85~GeV, together with the collected 
integrated luminosity of 57~pb$^{-1}$, 
lead to excellent discovery potential.
This paper describes a chargino and neutralino search using the above
data sample  
and an analysis which is improved relative to the one presented in  
a previous publication~\cite{LEP17-opal}.

Previous searches for       
charginos and neutralinos have been performed by OPAL using data 
collected near the Z peak 
(LEP1), at $\roots=$130--136~GeV~\cite{LEP15-opal}, 
at 161~GeV~\cite{LEP16-opal} and at 170--172~GeV~\cite{LEP17-opal}, and
by the other LEP collaborations~\cite{LEP15-chargino}~\cite{LEP17-chargino}. 

Charginos, $\chpm_{j}$, are the mass eigenstates
formed by the mixing of
the fields of the fermionic partners of the W boson (winos)
and those of the charged Higgs bosons (charged higgsinos).
Fermionic partners of the $\gamma$, Z,
and of the neutral Higgs bosons
mix to form mass eigenstates called neutralinos, $\nt_{i}$.
In each case, the index $j$
or $i$ is ordered by increasing mass.
R-parity~\cite{rpv} conservation is assumed;
therefore, the lightest supersymmetric particle (LSP)
is stable. The LSP is usually considered to be 
the lightest neutralino, $\nt_1$,
although it could be the scalar
neutrino, $\tilde{\nu}$, if it is sufficiently light.
The LSP is undetected due to its weakly interacting nature.
The present analysis is valid for either choice of the LSP.
In the Minimal Supersymmetric Standard Model (MSSM) 
there are two chargino mass eigenstates ($\ch_1$ and $\ch_2$) 
and four neutralino mass eigenstates ($\nt_1$, $\nt_2$, $\nt_3$ and $\nt_4$).

If charginos exist and are sufficiently light,
they are pair-produced
through a $\gamma$ or Z in the $s$-channel.
For the wino component there is an additional production process 
through scalar electron-neutrino ($\snu_{\mathrm{e}}$) 
exchange in the $t$-channel.
The production cross-section is large
unless the scalar neutrino (sneutrino) is light, in which case
the cross-section is reduced by destructive interference
between the $s$-channel $\mathrm e^+e^-$ annihilation to
Z or $\gamma$ and $t$-channel $\snu_{\mathrm{e}}$ 
exchange diagrams~\cite{Bartl, chargino-theory}.
The details of chargino decay depend on
the parameters of the mixing
and the masses of the scalar partners of the ordinary fermions.
The lightest chargino $\chp_1$ can decay
into $\nt_1 \ell^+ \nu$, or
$\nt_1 \mathrm{q} \mathrm{\overline{q}'}$,
via a W boson,
scalar lepton ($\sell$, $\snu$)
or scalar quark (squark, $\sq$).
In much of the MSSM parameter space, $\chp_1$
decays via a W boson are dominant. 
Due to the energy and momentum carried away
by the LSP (and possibly by neutrinos),
the experimental signature for $\chp_1 \chm_1$ events
is large missing energy and large missing momentum transverse
to the beam axis.
If the sneutrino is lighter than the chargino,
the two-body decay $\chp_1 \rightarrow \snu \ell^+$
dominates.  Special attention is paid to the case
$m_{\snu} \approx m_{\chpm_1}$ that would result in
two low-momentum charged leptons.

Neutralino pairs ($\nt_1 \nt_2$) can be produced through an
$s$-channel virtual Z, or by $t$-channel scalar 
electron (selectron, $\sele$)
exchange \cite{Ambrosanio}.
The MSSM prediction
for the $\nt_1 \nt_2$ production cross-section
can vary significantly
depending on the choice of MSSM parameters.
It is typically a fraction of a picobarn and generally much lower than
the cross-section for $\chm_1 \chp_1$ production.
The $\nt_2$ will decay into 
$\nt_1 \nunu$, $\nt_1 \ellell$ or
$\nt_1 {\mathrm q}\bar{\mathrm q}$,
through a Z$^{(*)}$ boson,
sneutrino, slepton, squark or
a neutral SUSY Higgs boson ($\mathrm{h}^0$ or $\mathrm{A}^0$)\@.
The decay via Z$^{(*)}$ is the dominant mode in most of 
the parameter space.
For the cases of $\nt_2 \ra \nt_1 \ellell$ or
$\nt_1 {\mathrm q}\bar{\mathrm q}$,
this leads to an experimental signature consisting either
of an acoplanar pair
of particles or jets, or a monojet if the two jets in
the final state have merged.
The radiative decay process
$\nt_2 \ra \nt_1 \gamma$ is also possible~\cite{raddecay} and
can dominate for some regions of the parameter space.

Motivated by Grand Unification and to simplify the physics interpretation,
the Constrained Minimal Supersymmetric
Standard Model (CMSSM)~\cite{Bartl,chargino-theory,Ambrosanio,Carena}
is used to guide the analysis but more
general cases are also studied.
In the CMSSM all the gauginos (SUSY partners of $U(1)_Y$, 
$SU(2)_L$ and $SU(3)_c$ gauge
bosons) are assumed to have a common mass, $m_{1/2}$, at the 
grand unified (GUT) mass scale,  
and all the sfermions (SUSY partners of quarks and leptons)
have a common mass, $m_0$, at the GUT mass scale.   

In the CMSSM analyses reported here all possible
cascade decay processes \cite{Ambrosanio,Carena} are taken into account. 
For example, if $\nt_2$ is lighter than $\chp_1$, 
the cascade decay of the chargino, 
$\chp_1 \ra \ffbar \nt_2 , (\nt_2 \ra  \ffbar \nt_1$,
or $ \nt_2 \ra  \gamma \nt_1 $), is  possible\footnote{
Pair production of $\nt_2 \nt_2$ is also possible,
but direct searches for this channel with the decays
$\nt_2 \rightarrow \nt_1 \mathrm{Z}^{(*)}$ would contribute negligibly
to the overall limits placed on the CMSSM parameter space, and
direct searches for this mode are not made in this analysis.
The process $\nt_2 \nt_2 \ra \gamma \nt_1 \gamma \nt_1$ is 
taken into account in the CMSSM limits calculations using
the experimental results of another OPAL analysis~\cite{gamgam}.
}\@.   
The production of $\nt_3$ is also taken into account in the analysis.   
The experimental signatures for the $\nt_1 \nt_3$ production 
are similar to those for $\nt_1 \nt_2$,
if $\nt_3$ decays into $\nt_1 \mathrm{Z}^{(*)}$, or
into $\nt_1 \mathrm{h}^0$, $\nt_1 \mathrm{A}^0$ or $\nt_1 \gamma$.
%

This paper is organised as follows.
The various event simulations which have been used are described in
Section 2. 
Analyses of the possible signal topologies
are discussed in Section 3 and
results and physics interpretations, both model independent and 
based on the CMSSM, are given in Section~4.

\section{The OPAL Detector and Event Simulation}
\subsection{The OPAL Detector}
%
The OPAL detector
is described in detail in \cite{OPAL-detector}\@.
It is a multipurpose apparatus
having nearly complete solid angle coverage\footnote{
A right-handed coordinate system is adopted,
where the $x$-axis points to the centre of the LEP ring,
and positive $z$ is along the electron beam direction.
The angles $\theta$ and $\phi$ are the polar and azimuthal angles,
respectively.}.
The central tracking system consists of a silicon microvertex
detector, a vertex drift chamber, a jet chamber and $z$-chambers.
In the range $|\cos\theta|<0.73$, 159 points can be measured in the
jet chamber along each track. At least 20 points on a track can be
obtained over 96\% of the full solid angle.
The whole tracking system is located
inside a 0.435~T axial magnetic field. 
A lead-glass electromagnetic (EM) calorimeter providing acceptance within
$|\cos\theta|<0.984$ together with
presamplers and time-of-flight scintillators
is located both outside the magnet coil and at the front of each endcap.
The magnet return yoke is instrumented for hadron calorimetry (HCAL),
giving a polar angle coverage of $|\cos\theta|<0.99$,
and is surrounded by external muon chambers.
The forward detectors (FD) and
silicon-tungsten calorimeters
(SW) located on both sides of the interaction point measure the luminosity
and complete the geometrical acceptance
down to 24~mrad in polar angle.
The small gap between the endcap EM calorimeter and FD
is filled by an additional electromagnetic calorimeter,
called the gamma-catcher (GC).

\subsection{Event Simulation}
%
The DFGT generator \cite{DFGT} 
is used to simulate signal events.
It includes spin correlations and
allows for a proper treatment of both the W boson and the Z boson 
width effects in the chargino and heavy neutralino decays.
The generator includes initial-state radiation and uses
the JETSET~7.4 package~\cite{PYTHIA} for the hadronisation
of the quark-antiquark
system in the hadronic decays of charginos and neutralinos.
SUSYGEN~\cite{SUSYGEN} is used to calculate 
the branching fractions for the CMSSM interpretation of the analysis. 
The most important parameters influencing the chargino detection efficiency
are the mass of the lightest chargino, $m_{\chp_1}$,
and the mass difference between the lightest chargino and the lightest
neutralino, $\Delta M_+ \equiv m_{\chp_1} - m_{\nt_1}$.
$\chp_1 \chm_1$ events are generated for 78 points 
in the ($m_{\chp_1}$,$\Delta M_+$) plane, for $m_{\chp_1}$ 
between 50~GeV and 90~GeV and $\Delta M_+$ between 3~GeV
and $m_{\chp_1}$. 
At each point 1000 events for the decay
$\chp_1 \ra \nt_1 {\mathrm W}^{*+} $
are generated.
For the two-body decays of the chargino ($\chp_1 \ra \snu \ell^+$),
43 points of 1000 events are generated 
in the ($m_{\chp_1}$,$m_{\snu}$) plane, for $m_{\chp_1}$ 
between 45~GeV and 90~GeV and $m_{\snu}$ between 1.5~GeV
and $m_{\chp_1}-35$~GeV.
For the $\nt_1 \nt_2$ production, $m_{\nt_2}$ and
$\Delta M_0 \equiv m_{\nt_2} - m_{\nt_1}$ are the main parameters
affecting the efficiency.
The $\nt_1 \nt_2$ events are generated in 75 points 
of the ($m_{\nt_2}$,$\Delta M_0$) plane, for $m_{\nt_1}+m_{\nt_2}$ 
between 100~GeV and 180~GeV and $\Delta M_0$ between 3~GeV
and $m_{\nt_2}$\@. 
For the cascade decay of $\ch_1$, 55 points are generated
varying $m_{\nt_2}, m_{\nt_1}$, and the branching fractions for  
$\chp_1 \ra \ffbar \nt_2$ and $\nt_2 \ra \nt_1 \gamma$\@.

The sources of background to the chargino and neutralino signals
are two-photon, lepton pairs, multihadronic and four-fermion
processes.
Two-photon processes are the most important background
for the case of small $\Delta M_+$ and small $\Delta M_0$ where
the visible energy and momentum transverse to
the beam direction for signal and two-photon events are comparatively small.
The Monte Carlo generators
PHOJET~\cite{PHOJET} (for $Q^2 < 4.5$~GeV$^2$) and 
HERWIG~\cite{HERWIG} (for $Q^2 \geq 4.5$~GeV$^2$)
are used to simulate hadronic events from two-photon
processes.
The Vermaseren~\cite{Vermaseren} program is used to
simulate leptonic two-photon processes  
($\ee\ee$, $\ee \mumu$ and $\ee \tautau$). 
Four-fermion processes 
are simulated using the grc4f~\cite{grc4f}
generator, which takes into account all interferences.   
The dominant contributions are
$\WW$, $\Wenu$, $\gamma^* \Z$ and $\Z\Z^{(*)}$ events, which have
topologies very similar to that of the signal.
Additional samples of $\ee\ee$, $\ee \mumu$ and $\ee \tautau$ processes 
which are not covered by the Vermaseren program
are generated using grc4f. 
Lepton pairs are generated using
the KORALZ~\cite{KORALZ} generator for
$\tau^+ \tau^- (\gamma)$ and $\mumu (\gamma)$ events,
and the BHWIDE~\cite{BHWIDE} program
for $\ee \ra \ee (\gamma) $ events.
Multihadronic ($\qq (\gamma)$) events are simulated using
PYTHIA~\cite{PYTHIA}. 

Generated signal and background events are processed
through the full simulation of the OPAL detector~\cite{GOPAL}
and the same event analysis chain was applied to the simulated events
as to the data.
%
\section{Data Analysis}
\label{analysis}

The analysis is performed on data
collected during the 1997 run of LEP
at $\roots = $181--184~GeV. 
The luminosity weighted average\footnote{Most of the data was taken at 
182.7~GeV,
3.3\% of the data was taken at 183.8~GeV,
6.1\% at 181.8~GeV  and 
0.6\% at less than 181.0~GeV.
The error of $\roots$ from the beam energy uncertainty 
is 0.03~GeV.} of $\roots$ is 182.7~GeV.
  
Data are used from runs in which all the subdetectors
relevant to this analysis were fully operational, corresponding
to an integrated luminosity of $56.75 \pm 0.27$~pb$^{-1}$.
The luminosity is measured using small-angle Bhabha scattering 
events detected in the silicon-tungsten calorimeter.

The following preselection cuts are applied to all data to select
well measured events:
(1) the number of charged tracks is required to be at least two; 
(2) the event transverse momentum relative to the beam direction
is required to be larger than 1.8~GeV\@;
(3) the total energy deposit
in each side of the SW, FD and GC detectors
has to be smaller than 2, 2 and 5~GeV, respectively;
(4) the visible invariant mass of the event has to exceed 2 GeV\@;
(5) the maximum EM cluster energy 
and the maximum charged track momentum has to be smaller than 
130\% of the beam energy to reject badly reconstructed events;
and (6) the absolute mean of the impact parameters (with respect to the
beam spot position) of the reconstructed charged tracks
are required to be smaller than 1~cm for the high multiplicity analyses
to reject background from beam-gas and beam-wall interactions.

\subsection{Detection of Charginos}

To obtain optimal performance the event sample is divided exclusively 
into three 
categories, motivated by the topologies expected from chargino
events. 
Separate analyses are applied to the preselected events in 
each category:
\begin{itemize}
\item[(A)] 
$N_{\mathrm ch} > 4$ and no isolated leptons,
where $N_{\mathrm ch}$ is the 
number of charged tracks: 
when both $\chp_1$ and $\chm_1$ decay
hadronically, signal events tend to fall 
into this category for modest and large values of 
$\Delta M_+ (\equiv m_{\chp_1} - m_{\nt_1})$\@.
\item[(B)] 
$N_{\mathrm ch} > 4$ and at least one isolated lepton:
if only one of the $\chpm_1$ decays leptonically,
signal events tend to fall 
into this category.
\item[(C)] 
$N_{\mathrm ch} \leq 4$:
events tend to fall into this category if $\Delta M_+$ is small or if
both charginos decay leptonically.
\end{itemize}
The isolated lepton selection criteria
are similar to those described in Ref.~\cite{LEP17-opal}. 
Electrons are selected if they satisfy either of the two
identification methods described in~\cite{e1, e2} and muons are
identified using three methods~\cite{mu1, mu2, mu3}.
The momentum of the electron or muon candidate
is required to be larger than 2~GeV.
In order to identify a tau lepton,
events are reconstructed using 
the Durham jet algorithm with a jet resolution parameter of 3~GeV$^2$\@.
A reconstructed jet is identified as a tau decay if
there are only one or three charged tracks in the jet, if
the momentum sum of the charged tracks is larger than 2.0~GeV,
if the invariant mass of the charged particles in the jet is
smaller than 1.5~GeV and if the invariant mass of the jet is 
smaller than 2.0~GeV.
The identified lepton is defined to be isolated if the
energy within a cone of half-angle $20\degree$ around
the electron, muon or tau candidate
is less than 2~GeV.

The fraction of $\chp_1 \chm_1$ events falling into
category (A) is about 35-50\% for most $\Delta M_+$ values.
This fraction drops to less than 15\% if $\Delta M_+$ is smaller than 5~GeV,
since the average charged track multiplicity of the events is smaller.
Similarly, 
the fraction of events falling into category (B) is also about 35-50\% 
for most $\Delta M_+$ values and is less than 10\% if 
$\Delta M_+$ is smaller than 5~GeV.
When $\Delta M_+$ is smaller than 10~GeV,
the fraction of events falling into category (C) is greater than about 50\%\@.
If $\Delta M_+$ is larger than 20~GeV, 
this fraction is about 10\%\@.

Since the chargino event topology mainly depends on the 
difference between the chargino mass and the lightest neutralino
mass, different selection criteria are applied to four 
$\Delta M_+$ regions:  
\begin{itemize}
\item[(I)]
$\Delta M_+ \leq 10$~GeV,
\item[(II)]
10~GeV $< \Delta M_+ \leq m_{\chp_1}/2$,
\item[(III)]
$ m_{\chp_1}/2 < \Delta M_+ \leq m_{\chp_1}-20$~GeV,
\item[(IV)]
$ m_{\chp_1}-20$~GeV$ < \Delta M_+ \leq m_{\chp_1}$. 
\end{itemize}
In region I, the main background comes from
two-photon processes.
Background from four-fermion processes (mainly $\WW$) can be safely rejected 
without sacrificing  signal detection efficiency.
In regions II and III, the main background comes from 
four-fermion processes ($\WW$, single W and $\gamma^* Z^{(*)}$).
In these regions the background level is modest.
In region IV the $\WW$ background is large and dominant.
Since the $\WW$ background is very severe in the region of 
$\Delta M_+ > 85$~GeV where the chargino decays via an on-mass-shell
W-boson, a special analysis (allowing 
relatively high background and demanding 
higher signal efficiency)  
is applied to improve the sensitivity
to the chargino signal.
Overlap between this analysis and the region IV standard analysis 
is avoided by selecting the analysis which minimises the expected
cross-section limit calculated with only the expected background number.
This generally results in the special analysis being applied
for $m_{\chp_1} > 85$~GeV and $\Delta M_+ \gsim 85$~GeV. 
If $m_{\snu}$ is lighter than $\chp_1$,
the two-body decays of the chargino ($\chp_1 \ra \snu \ell^+$)
may dominate over the three-body decays via a virtual W.
A special analysis, optimised for this case, is also performed
in category (C).

For each region a single set of cut values is determined which
minimises
the expected limit on the signal cross-section at 95\% C.L. using 
Ref.~\cite{PDG96}. For this procedure, it is 
assumed that the distribution of observed candidates
arises from the expected number of background events
and therefore the choice of cuts is 
independent of the number
of candidates actually observed.

The efficiency for an arbitrary choice of $m_{\chp_1}$
and $m_{\nt_1}$ is obtained by interpolation using a polynomial fit to
the efficiencies determined from the Monte Carlo 
for each category in each $\Delta M_+$ region.

\subsubsection{Analysis (A) (\boldmath$N_{\mathrm ch} > 4$
without isolated leptons)}

\begin{table}[htb]
  \centering
  \begin{tabular}{|l||r|r|r|r|}
    \hline
    Region  &  I~~ &  II~ &  III & IV~  \\
    \hline
    \hline
    $E_{\mathrm fwd}/E_{\mathrm vis}$
    &  \multicolumn{4}{|c|}{ $<0.2$ } \\
    \hline
    $|\cos \theta_{\mathrm miss}|$
    & \multicolumn{4}{|c|}{ $<0.85$ } \\
    \hline
    $|P_z|$~GeV & \multicolumn{4}{|c|}{ $<30$ } \\
    $|P_z|/E_{\mathrm vis}$ & \multicolumn{4}{|c|}{ $<0.7$ } \\
    \hline
    $P_{\mathrm t}$~GeV & \multicolumn{4}{|c|}{ $>6$ } \\
    \cline{2-5}
    $P_{\mathrm t}^{\mathrm HCAL}$~GeV & [6,30] &
    \multicolumn{3}{|c|}{ $>6$ } \\
    \hline
    $N_{\mathrm jet}$ & 
    \multicolumn{2}{|c|}{ [2,5] } & \multicolumn{2}{|c|}{ [3,5] } \\
    \hline
    $|\cos \theta_{\mathrm j}|$
    & \multicolumn{4}{|c|}{ $<0.90$ }  \\
    \hline
    $\phi_{\mathrm acop}^{\circ}$&\multicolumn{2}{|c|}{$>15(10^*)$}&
    \multicolumn{2}{|c|}{$>15$} \\
    \hline
    if $M_{\mathrm vis} >90$,  & \multicolumn{2}{|c|}{ } 
    &  \multicolumn{2}{|c|}{ } \\
    $p_{\mathrm max}$~GeV &\multicolumn{2}{|c|}{--}  
    & \multicolumn{2}{|c|}{$<20$} \\
    \hline
    $M_{\mathrm vis}$~GeV&$<60$&$[5,80]$&$[5,130]$&$[5,150]$ \\
    \hline
    if $N_{\mathrm \ell'} = 1$  &    &    &    & \\
    $M_{\mathrm had'}$~GeV & $<30$ & $<50$ & $<60$ & $\{60,90\}$   \\
    \cline{2-5}
    $E_{\ell'}$~GeV        &\multicolumn{3}{|c|}{$<30$} & $<35$ \\
    \hline
    $E_1$~GeV & --  & [2,35] & [2,50] & [2,55]  \\
    \hline
    $E_2$~GeV & --  & [2,25] & [2,50] & [2,55]  \\
    \hline
    if $N_{\mathrm jet} = 3$  & \multicolumn{2}{|c|}{ } 
    & \multicolumn{2}{|c|}{ } \\
    $|P_z|$~GeV               &\multicolumn{2}{|c|}{--} 
    &\multicolumn{2}{|c|}{$<10^{**}$} \\
    \hline
    \hline
    background &   &   &   &    \\
    \hline
    $\gamma \gamma$    & 0.23 & 0.11 & 0.11 & 0.11 \\
    \hline
    $\ellell (\gamma)$ & 0.02 & 0.02 & 0.01 & 0.01 \\
    \hline
    $\qq (\gamma)$     & 0.10 & 0.11 & 0.40 & 0.82 \\
    \hline
    4f                 & 0.67 & 1.09 & 5.89 & 8.43 \\
    \hline
    \hline
    total bkg.
    & $ 1.01\pm0.17 $ & $1.33\pm0.16$ & $6.42\pm0.29$ & $9.38\pm0.34$ \\
    \hline
    \hline
    observed & 1 & 1 & 4 & 7   \\
    \hline
  \end{tabular}
  \caption[]{
    \protect{\parbox[t]{12cm}{
        The list of selection criteria for category (A).
        The numbers of background events expected for the integrated luminosity
        of $\lumi$~pb$^{-1}$, for various Standard Model processes, 
        and the total
        number of background events expected as well as the observed number of
        events in each $\Delta M_+$ region are also listed.
        If $N_{\mathrm  jet} = 2$, $\phi_{\mathrm acop}$ is
        required to be larger than $10^{\circ}$ (as indicated by *).
        $N_{\mathrm jet}=$ [2, 5] means that 
        $2 \leq N_{\mathrm jet} \leq 5$. 
        $M_{vis}=\{60,90\}$ means that $M_{vis}<60$ or $M_{vis}>90$~GeV. 
        If $N_{\mathrm  jet} = 3$ and $70 < M_{\mathrm vis} < 95$~GeV
        in region III and IV, $|P_z|$ is required to be smaller 
        than $10$~GeV (as indicated by **).
        The errors in the total background include only the statistical
        errors of simulated background events. 
        }} }
  \label{tab:cutAch}
\end{table}

For modest and large values of $\Delta M_+$,
if both $\chp_1$ and $\chm_1$ decay
hadronically, signal events tend to fall 
into category (A).
The variables used in the selection criteria and their cut values 
optimised in each $\Delta M_+$ region 
are listed in Table~\ref{tab:cutAch}.

After the preselection, the cuts on $E_{\mathrm fwd}/E_{\mathrm vis}$, 
$|\cos \theta_{\mathrm miss} |$, $|P_z|$ and $|P_z|/E_{\mathrm vis}$, 
where $E_{\mathrm vis}$ is the total visible energy of the event, 
$E_{\mathrm fwd}$ is the visible energy in the region of 
$| \cos \theta |>0.9$, $\theta_{\mathrm miss}$ is the polar angle
of the missing momentum and $P_z$ is the visible momentum
along the beam axis, are applied to reduce background from two-photon 
and $\mathrm Z^{0}$ radiative return processes. 
Most of the remaining background from two-photon processes is rejected 
by the cuts on $P_{\mathrm t}$ and $P_{\mathrm t}^{\mathrm HCAL}$, 
the transverse momentum of the event measured without using
the hadron calorimeter and using the hadron calorimeter, respectively.
In region I, the upper cut on  $P_{\mathrm t}^{\mathrm HCAL}$ 
is determined from the range of the signal $P_{\mathrm t}^{\mathrm
  HCAL}$ distribution; 
it reduces the background from $\WW$ and $\Wenu$ events.
 
The number of jets ($N_{\mathrm jet}$) is reconstructed 
using the Durham jet algorithm
with jet resolution parameter $y_{\mathrm cut} = 0.005$.
With the $N_{\mathrm jet}$ cut, small invariant mass monojet events from 
the process $\gamma^{*} {\mathrm Z}^{(*)} \ra \qq \nunu$ are rejected 
for regions I and II, and background from 
$\qq (\gamma)$ and single W events is reduced in
regions III and IV.

\begin{figure} \centering
\begin{minipage}{16.0cm}
\begin{center}\mbox{
\epsfig{file=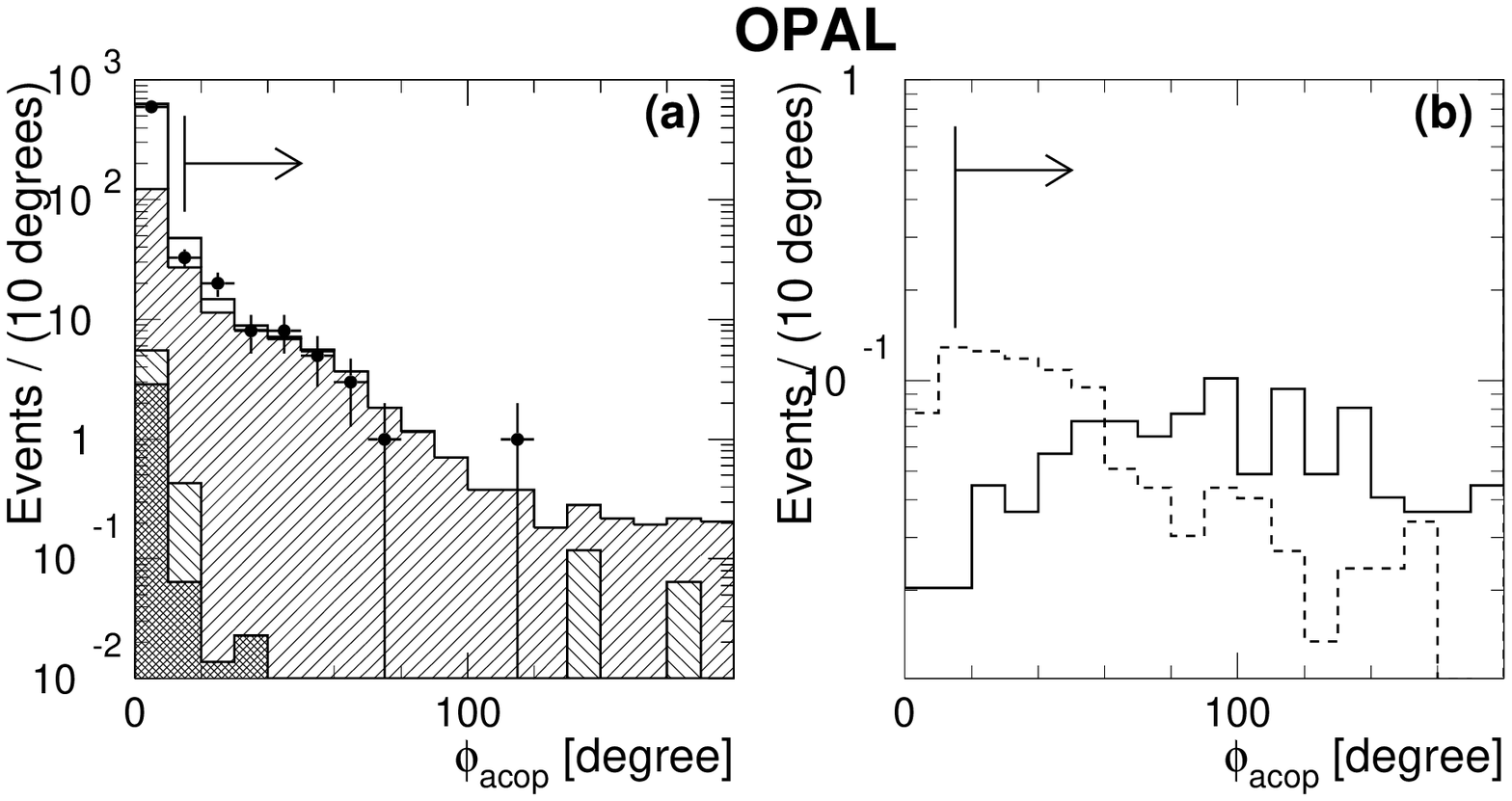,width=16.0cm}
}\end{center}
\vspace{-9mm}
\caption[]
{The distributions of the acoplanarity angle, $\phiacop$, 
in analysis (A) region II.
In (a) are shown the data distribution (dark circles)
and the predicted contributions
from background processes: 
dilepton events (double hatched area), 
two-photon processes (negative slope hatched area), 
four-fermion processes (including W-pair events) 
(positive slope hatched area),
and multihadronic events (open area). In each case the distribution
has been normalised to $\lumi$~pb$^{-1}$. 
In (b) predictions from
simulated chargino events are shown for $m_{\ch_1}=90$~GeV and 
$m_{\nt_1}=70$~GeV (solid line histogram) and for
$m_{\ch_1}=90$~GeV and $m_{\nt_1}=45$~GeV (dashed line histogram).
The normalisations of the signal distributions are arbitrary. 
The arrows shown indicate the selection criteria.
}
\label{fig:cataacop}
\end{minipage}
\end{figure}

After the above cuts, to calculate the acoplanarity angle 
($\phi_{\mathrm acop}$), 
the events are forced into two jets, again using the Durham jet algorithm.
The polar angle of each jet, $\theta_{\mathrm j}$ (j=1,2), 
is required to be far from the beam axis. 
This cut ensures a good measurement of $\phi_{\mathrm acop}$
and further reduces the background from the $\qq (\gamma)$ and
two-photon processes.  
To select signals with a large amount of missing momentum due to the 
two invisible neutralinos and to reduce $\qq (\gamma)$
background, the acoplanarity angle is required to be large.
The acoplanarity angle distributions for region II 
are shown in Fig.~\ref{fig:cataacop} before application of this cut.

\begin{figure} \centering
\begin{minipage}{16.0cm}
\begin{center}\mbox{
\epsfig{file=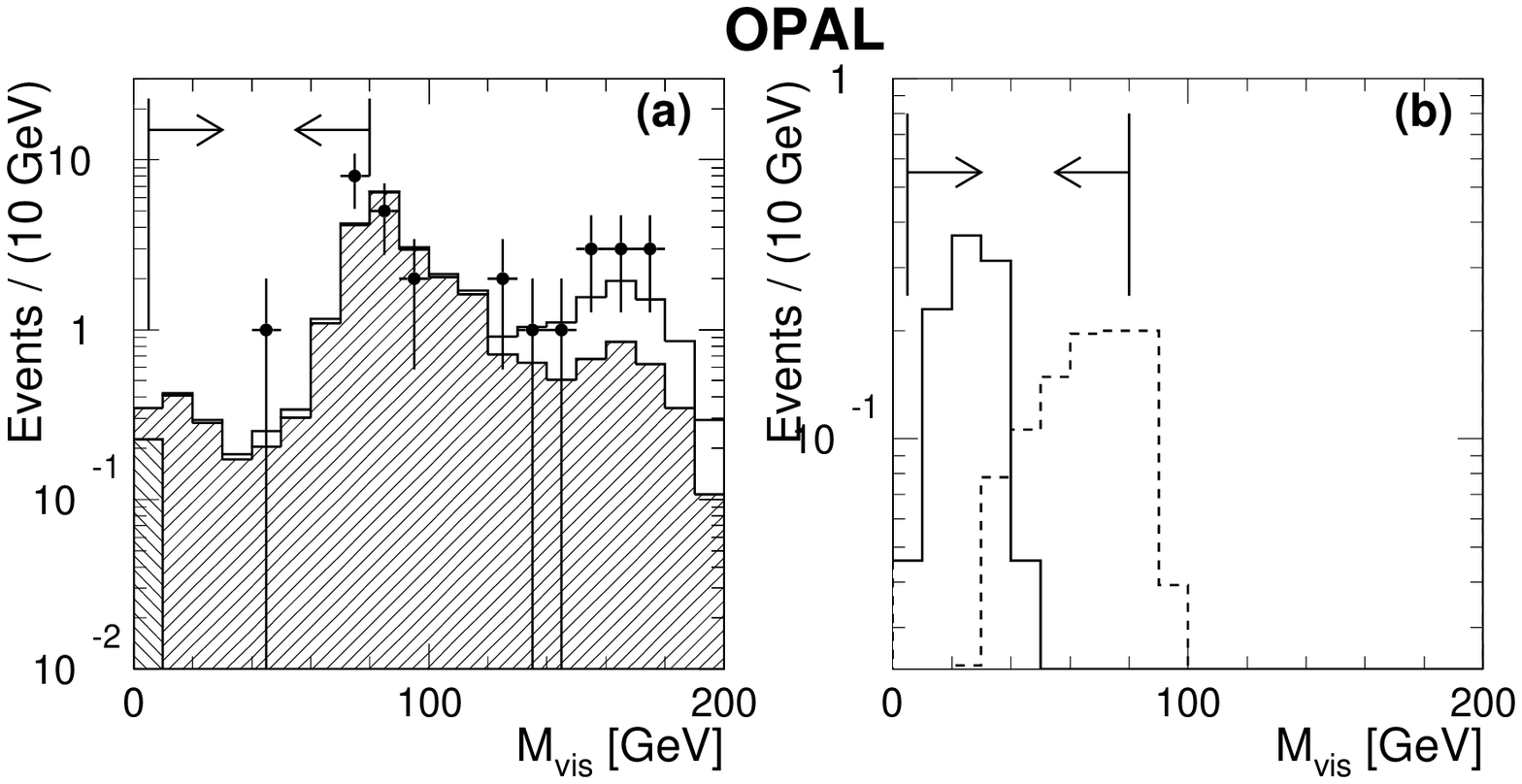,width=16.0cm}
}\end{center}
\vspace{-9mm}
\caption[]
{The distributions of $\Mvis$ in analysis (A)
for data and simulated background events are 
shown in (a) after the $\phiacop$ cut for region II.
The background sources are shaded as in Fig.1. 
Distributions of chargino signal events are shown in
(b) for $m_{\ch_1}=90$~GeV and  $m_{\nt_1}=70$~GeV (solid line) and for
$m_{\ch_1}=90$~GeV and  $m_{\nt_1}=45$~GeV (dashed line). 
The normalisations of the signal distributions are arbitrary. 
The arrows shown indicate the selection criteria.
}
\label{fig:catamvis}
\end{minipage}
\end{figure}

For events with an observed invariant mass ($M_{\mathrm vis}$)
greater than 90~GeV, a cut is applied on the maximum track 
momentum ($p_{\mathrm max}$) in regions III and IV.
This cut reduces $\WW \ra \ell \nu \qq'$ events where the lepton $\ell$
overlaps with a hadronic jet.
The $M_{\mathrm vis}$ cut is optimised for each $\Delta M_+$ region 
as shown in Table~\ref{tab:cutAch}.  
The distributions of the observed invariant mass
are shown in Fig.~\ref{fig:catamvis} for region II,
after the $\phiacop$ cut.

If a lepton ($\ell'$) is found with an algorithm based on the
looser isolation condition described in Ref.~\cite{LEP17-opal},
the lepton energy $E_{\ell'}$ and 
the invariant mass calculated without the lepton,
$M_{\mathrm had'}$, must be smaller than the values expected for 
$\WW \ra \ell \nu \qq'$ events. 

The backgrounds from $\WW$ events and single W events 
are efficiently suppressed by requiring that 
the highest energy ($E_1$) and the second highest energy ($E_2$) 
of the jets (reconstructed with $y_{\mathrm cut} = 0.005$)
are smaller than the typical jet energy
expected for the $\WW$ background.
In region III and IV three-jet events with 
$|P_Z| < 10$~GeV are rejected if $\Mvis$ is close to the W mass.
This cut reduces the $\WW \ra \tau \nu \qq'$ background with
small $\tau$ decay product energy.

The numbers of background events expected from the four different sources, for 
each $\Delta M_+$ region, are given in Table~\ref{tab:cutAch}.
Typical detection efficiencies for 
$\chp_1 \chm_1$ events are shown in Fig.~\ref{effi}a\@.

A special analysis is applied 
in the region of $\Delta M_+ \gsim 85$~GeV, since
the event topology of the signal is very similar to 
that of $\WW \ra$~4~jets.
This similarity is due to the small missing momentum 
taken by the low mass neutralinos.
After selecting well contained events 
with the cuts $|\cos \theta_{\mathrm miss}|<0.95$,
$E_{\mathrm fwd}/E_{\mathrm vis}<0.25$ and $|P_z|<30$~GeV,
multijet events with large visible energy are selected  
($N_{\mathrm jet} \geq 4$ and $100< E_{\mathrm vis} < 170$~GeV).
To select a clear 4-jet topology, $y_{34} \geq 0.0075$,
$y_{23} \geq 0.04$ and $y_{45} \leq 0.0015$ are required, where
$y_{\{n\}\{n+1\}}$ is defined as the minimum $y_{\mathrm cut}$ value
at which the event is reconstructed as having $n$-jets. 
If there is a jet which consists of a single $\gamma$ with energy greater than 
50~GeV, events are considered to be $\gamma \qq g$ and are rejected.
The maximum track momentum is required to be smaller than 40~GeV
to reduce background from 
$\WW \ra \ell \nu \qq'$ events where $\ell$ overlaps
with a hadronic jet. 
The number of selected events is 29, while  
the number expected from background processes 
is 24.7$\pm$0.5 (0.0 from $\gamma \gamma$,
0.0 from $\ellell (\gamma)$, 6.4 from $\qq (\gamma)$ and 
18.3 from four fermion final states).
The signal efficiency is 26--33\% for  
$m_{\chp_1} \geq 85$~GeV and $\Delta M_+ \geq 85$~GeV.

\subsubsection{Analysis (B) (\boldmath$N_{\mathrm ch} > 4$ 
with isolated leptons)}

\begin{table}[htb]
\centering
\begin{tabular}{|l||r|r|r|r|}
\hline
Region  & I~~ & II~ & III &  IV~  \\
\hline
\hline
$E_{\mathrm fwd}/E_{\mathrm vis}$ 
& $<0.15$ & $<0.2$ & \multicolumn{2}{|c|}{$<0.3$}  \\ 
\hline
$|\cos \theta_{\mathrm miss}|$ 
&  \multicolumn{4}{|c|}{$<0.9$}  \\ 
\hline
$P_{\mathrm t}$~GeV & $>4$ & $>5$ & $>7$ & $>8$  \\ 
\cline{2-5}  
$P_{\mathrm t}^{\mathrm HCAL}$~GeV & $>4$ & $>5$ & $>7$ & $>8$  \\  
\hline
$|\cos \theta_{\mathrm j}|$ & \multicolumn{4}{|c|}{$<0.95$} \\ 
\hline
$\phi_{\mathrm acop}^{\circ}$ & \multicolumn{2}{|c|}{$>20$} 
                              & \multicolumn{2}{|c|}{$>15$} \\
\hline
$M_{\mathrm had}$~GeV & $<30$ & $<60$ & [5,60] & [5,65]  \\  
\hline
$p_{\ell}$~GeV & $<20$ & \multicolumn{2}{|c|}{$<45^{*}$} & [4,50] \\  
\hline
$M_{\mathrm vis}$~GeV & $<50$ & \multicolumn{2}{|c|}{$<80$} & $<80^{**}$  \\  
\hline
\hline
background &   &   &   &    \\  
\hline
$\gamma \gamma$    & 0.84 & 0.06 & 0.06 & 0.06  \\ 
\hline
$\ellell (\gamma)$ & 0.02 & 0.10 & 0.13 & 0.18  \\  
\hline
$\qq (\gamma)$     & 0.00 & 0.02 & 0.11 & 0.11  \\  
\hline
4f                 & 0.33 & 1.18 & 1.27 & 3.33  \\  
\hline
\hline
total bkg       
& 1.20$\pm$0.49 & 1.36$\pm$0.13 & 1.56$\pm$0.14 & 3.68$\pm$0.21  \\  
\hline
\hline
observed       & 0  &  1  &  1  &  3  \\  
\hline
\end{tabular}
\caption[]{
  \protect{\parbox[t]{12cm}{
The list of selection criteria for category (B).
The $p_{\ell}$ cut for regions II and III (as indicated by
$*$) are a function of $M_{\mathrm had}$
($p_{\ell}< 4.5 M_{\mathrm had}$ if $M_{\mathrm had} < 10$~GeV,
$p_{\ell} \leq 45$~GeV  if $10 \leq M_{\mathrm had} \leq 50$~GeV,
$p_{\ell} \leq -4.5 (M_{\mathrm had}-60$~GeV)
if $M_{\mathrm had} > 50$~GeV).
The  $M_{\mathrm vis}$ cut in region IV 
(as indicated by $**$) is only applied if
$M_{\mathrm had}> 55$~GeV.
The numbers of background events expected for the integrated luminosity 
of $\lumi$~pb$^{-1}$ for various Standard Model processes and the total
number of background events expected as well as the observed number of 
events in each $\Delta M_+$ region are also listed.
The errors in the total background include only the statistical error
of simulated background events.
}} 
}
\label{tab:cutBch}
\end{table}

\begin{figure} \centering
\begin{minipage}{16.0cm}
\begin{center}\mbox{
\epsfig{file=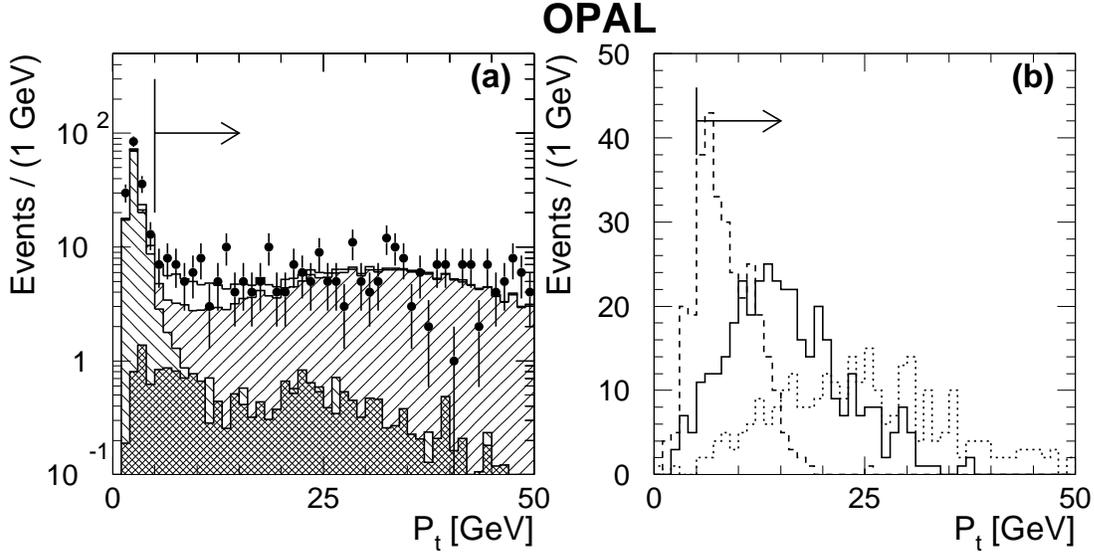,width=16.0cm}
}\end{center}
\vspace{-9mm}
\caption[]
{
The distributions of $P_t$ in analysis (B) region II
after the $|\cos \theta_{\mathrm miss}|$ cut.
Data and expected background contributions are shown in (a).
The background sources are shaded as in Fig.1. 
The distributions of the signal for simulated chargino events with
$m_{\chp_1} = 90$~GeV and $m_{\nt_1} = 70$~GeV (solid histogram),
with $m_{\chp_1} = 90$~GeV and $m_{\nt_1} = 80$~GeV (dashed histogram) and
with $m_{\chp_1} = 90$~GeV and $m_{\nt_1} = 45$~GeV (dotted histogram)
are shown in (b). 
The normalisations of the signal distributions are arbitrary. 
The arrows shown indicate the selection criteria.
}
\label{figb1}
\end{minipage}
\end{figure}

$\chp_1 \chm_1$ events in which one of the $\chpm_1$ decays
leptonically tend to fall into category (B).
The variables used in the selection 
and their cut values in each $\Delta M_+$ region  
are listed in Table~\ref{tab:cutBch}.
The signal events are selected with the following criteria:

Cuts on $|\cos \theta_{\mathrm miss}|$ and 
$E_{\mathrm fwd}/E_{\mathrm vis}$ are made to 
reject background from two-photon processes and 
$\qq (\gamma)$ events.
Large transverse momenta  
are then required to further reduce the contribution from 
two-photon events.
The distributions of $P_{\mathrm t}$ are shown in Fig.~\ref{figb1} 
after the $|\cos \theta_{\mathrm miss}|$ cut.
To calculate the acoplanarity angle,   
events are forced into two jets.
A large acoplanarity angle of the two jets is then required 
to further suppress the two-photon background.

\begin{figure} \centering
\begin{minipage}{16.0cm}
\begin{center}\mbox{
\epsfig{file=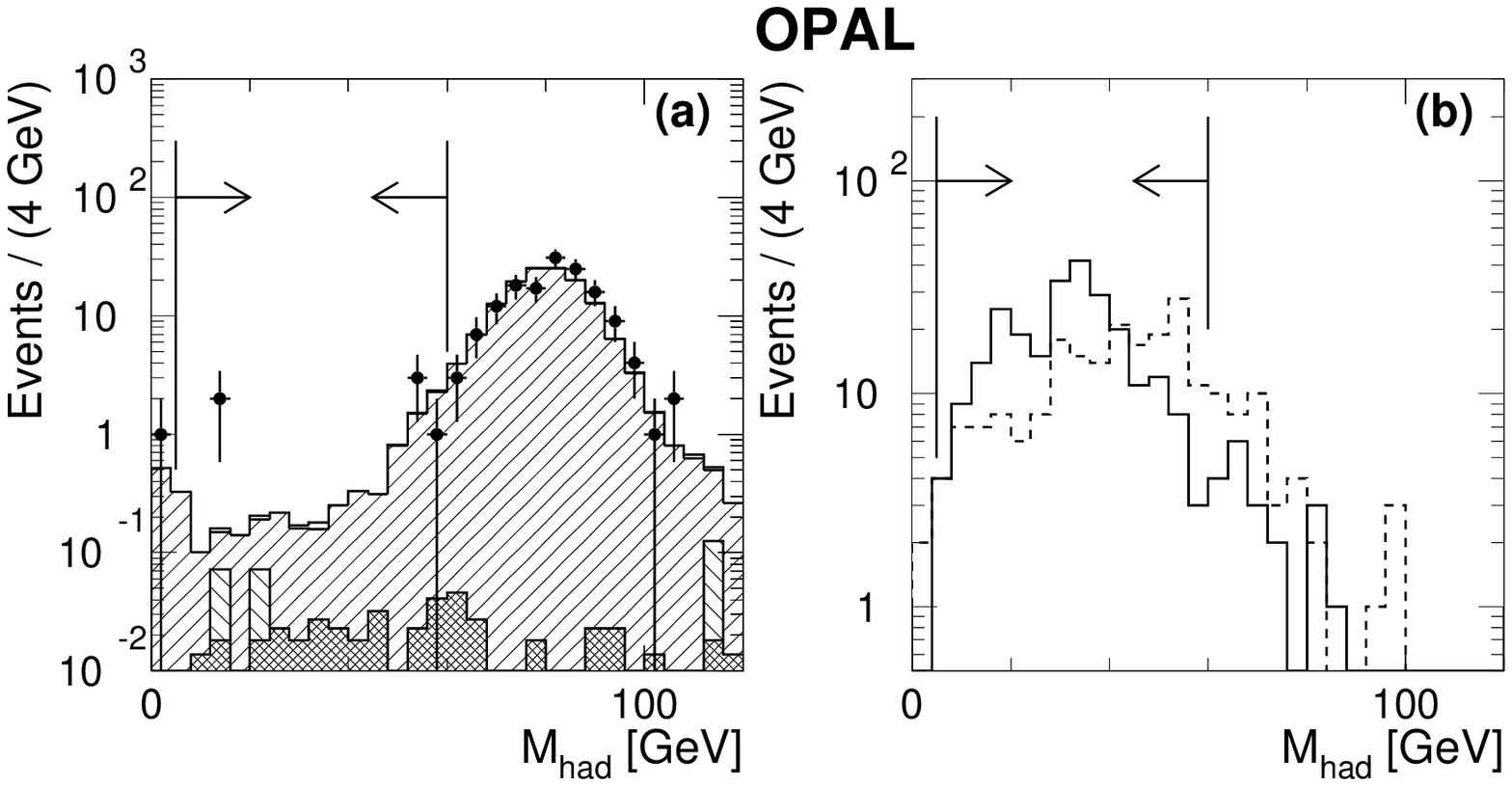,width=16.0cm}
}\end{center}
\vspace{-9mm}
\caption[]
{
$M_{\mathrm had}$ distributions in analysis (B) region III after 
the $\phiacop$ cut.
Figure (a) shows the data and the Monte Carlo prediction for 
background processes.
The background sources are shaded as in  Fig.~\ref{fig:cataacop}.
The distributions of the signal for simulated chargino events with
$m_{\chp_1} = 90$~GeV and $m_{\nt_1} = 45$~GeV (solid histogram) and
with $m_{\chp_1} = 90$~GeV and $m_{\nt_1} = 20$~GeV (dashed histogram)
are shown in (b). 
The normalisations of the signal distributions are arbitrary. 
The arrows shown indicate the selection criteria.
}
\label{figb3}
\end{minipage}
\end{figure}

In order to reject $\WW \ra \ell \nu \qq'$ background, 
various cuts are applied:
the momentum of isolated leptons should be small enough;
the invariant mass of the event calculated excluding the 
highest momentum isolated lepton ($M_{\mathrm had}$)
is required to be smaller than the W mass; 
in the small $\Delta M_+$ region the upper cut value of 
$M_{\mathrm had}$ is reduced, since signal events are concentrated
only in the small mass region. 
The distribution of $M_{\mathrm had}$ 
after  the $\phi_{\mathrm acop}$ cut is shown in Fig.~\ref{figb3} for
region III.
As shown in this figure, most of the $\WW$ background events are rejected.
Futhermore, We$\nu$ candidate events in which 
a fake lepton is found in the W$\ra \qq'$ decay
are further reduced by the $M_{\mathrm vis}$ cut.
Typical detection efficiencies for $\chp_1 \chm_1$ events 
are shown in Fig.~\ref{effi}a\@. 

A special analysis is applied 
in the region of $\Delta M_+ \gsim 85$~GeV due to the 
large $\WW$ background.  
The selection criteria are identical to those in region IV 
up to the $\phi_{\mathrm acop}$ cut as listed in 
the upper part of Table~\ref{tab:cutBch}.
To reject some
$\WW \ra \ell \nu \qq'$ events, while keeping a good 
signal efficiency, 
$M_{\mathrm had}$ is required to be between 50 and 90~GeV, and
$M_{\mathrm vis}$ to be between 80 and 130~GeV.
The invariant mass of the missing four-momentum and the 
four-momentum of the highest energy isolated lepton 
must lie between 90 and 130~GeV.
The number of observed events with these criteria is 28, while  
the number of expected background is 31.8$\pm$0.6, almost
all from four-fermion processes. 
The signal efficiency is 30--41\% for  
$m_{\chp_1} \geq 85$~GeV and $\Delta M_+ \geq 85$~GeV.

\subsubsection{Analysis (C) (\boldmath$N_{\mathrm ch} \leq 4$)}

Events in which both charginos decay leptonically 
as well as a large fraction of events for small $\Delta M_+$
tend to fall into category (C).
This analysis is especially important for the region of 
$\Delta M_+ \leq 5$~GeV. Because the background varies significantly
with $\Delta M_+$ in region I, this region has been split into 2 
sub-regions (a,b).
The cut variables and the cut values for each region and sub-region
are listed in Table~\ref{tab:cutCch}.

\begin{table}[htb]
\centering
\begin{tabular}{|l||r|r|r|r|r|}
\hline
Region  & \multicolumn{2}{|c|}{I} & II~ & III &  IV~  \\ \hline
Sub-region                    &
\multicolumn{1}{|c|}{a} & \multicolumn{1}{|c|}{b} &     &     &       \\
\hline
$N_{\mathrm ch}$              & \multicolumn{5}{|c|}{[2,4]}  \\ 
$\Sigma Q_{\mathrm i}$        & \multicolumn{5}{|c|}{0}  \\ 
\hline
$E_{\mathrm j}$~GeV           & \multicolumn{5}{|c|}{$>1.5$}  \\
$N_{\mathrm ch,j}$            & \multicolumn{5}{|c|}{$\geq 1$} \\
$|Q_{\mathrm j}|$             & \multicolumn{5}{|c|}{$\leq 1$} \\
\hline
for $\phi_{\mathrm acop}^{\circ}$ 
                              & \multicolumn{2}{|c|}{$\leq 70$} & \multicolumn{3}{|c|}{--} \\
$a_{\mathrm t}/E_{\mathrm beam}$ 
                              & \multicolumn{2}{|c|}{$>0.030$} & \multicolumn{3}{|c|}{--}  \\  
$P_{\mathrm t}/E_{\mathrm beam}$ 
                              & \multicolumn{2}{|c|}{$>0.030$} & \multicolumn{3}{|c|}{--} \\  
$|\cos \theta_a|$ 
                              & \multicolumn{2}{|c|}{$<0.975$} & \multicolumn{3}{|c|}{--} \\  
\hline
for $\phi_{\mathrm acop}^{\circ}$ & \multicolumn{2}{|c|}{$>70$} & \multicolumn{3}{|c|}{--} \\
\cline{2-6}
$P_{\mathrm t}/E_{\mathrm beam}$ 
                              & $>0.040$ & $>0.050$ & $>0.075$ & $>0.095$ & $>0.100$ \\  
\cline{2-6}
$|\cos \theta_{\mathrm miss}|$ 
                             & \multicolumn{3}{|c|}{$<0.90$} 
& \multicolumn{2}{|c|}{$<0.97$} \\  
\hline
$\Delta \phi _{(\vec P_{\mathrm miss}, \mu)}$~rad  
                              & \multicolumn{5}{|c|}{$>1.0$} \\ 
\hline
$|\cos \theta_{\mathrm j}|$   & \multicolumn{2}{|c|}{$<0.85$} & $<0.95$ &
\multicolumn{2}{|c|}{$<0.97$} \\  
\hline
$\phi_{\mathrm acop}^{\circ}$ & \multicolumn{2}{|c|}{[20,150]} & 
\multicolumn{3}{|c|}{$>$20}  \\
\hline
$M_{\mathrm vis}$~GeV         & $<$10.0 & $<$15.0 & $<$30.0 &
$<$40.0 & $<$60.0  \\  
\hline
$E_1 / E_{\mathrm beam}$      & $<0.15$ & $<0.22$ & $<0.30$ &
\multicolumn{2}{|c|}{$<1.00$}  \\  
\hline
\hline
background &  &  &  &  &  \\  
\hline
$\gamma \gamma$     & 1.99 & 0.68 & 2.04 & 1.53 & 1.31  \\  
\hline
$\ellell (\gamma)$  & 0.02 & 0.03 & 0.24 & 0.83 & 1.50  \\  
\hline
$\qq (\gamma)$      & 0.00 & 0.00 & 0.01 & 0.01 & 0.01 \\  
\hline
4f                  & 0.33 & 0.50 & 2.47 & 14.70 & 27.04 \\  
\hline
\hline
total bkg.    &  2.33$\pm$0.36& 1.21$\pm$0.21& 4.77$\pm$0.38
              & 17.08$\pm$0.48 & 29.86$\pm$0.55 \\     
\hline
\hline
observed       & 2  & 0  &  3  &  8  &  21  \\  
\hline
\end{tabular}          
\caption[]{
  \protect{\parbox[t]{12cm}{
The list of selection criteria for category (C).
Selection I(a) is specially optimised for the region of $\Delta M_+ \leq$ 5~GeV\@.  
The numbers of background events expected for the integrated luminosity 
of $\lumi$~pb$^{-1}$ for various Standard Model processes and the total
number of background events expected as well as the observed number of 
events in each $\Delta M_+$ region are also listed.
The errors in the total background include only the statistical error
of simulated background events.
}} }
\label{tab:cutCch}
\end{table}

\begin{figure} \centering
\begin{minipage}{16.0cm}
\begin{center}\mbox{
\epsfig{file=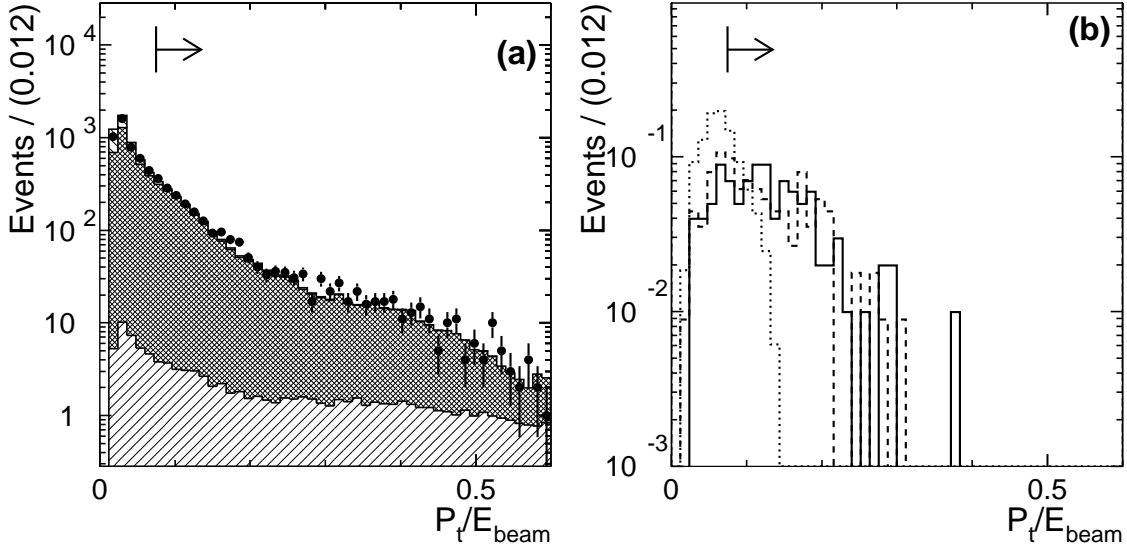,width=16.0cm}
}\end{center}
\vspace{-9mm}
\caption[]
{The distributions of $P_t/E_{\mathrm beam}$ 
after the cut on the jet charge for events in analysis (C) region II.
Data and background contributions are shown in (a). The
background sources are shaded as in Fig.~\ref{fig:cataacop}.
In (b) predictions from the simulation for chargino
and neutralino events are shown:
$m_{\chp_1}=90$~GeV and $m_{\nt_1}=70$~GeV (solid line), 
$m_{\chp_1}=90$~GeV and $m_{\nt_1}=45$~GeV (dashed line) and
$m_{\nt_2}=95$~GeV and $m_{\nt_1}=85$~GeV (dotted line).
The normalisations of the signal distributions are arbitrary.
The arrows shown indicate the selection criteria.
}
\label{catc1}
\end{minipage}
\end{figure}

The net charge of the event must be zero to reject poorly reconstructed events.
Since the signal is expected to have a two lepton or two jet topology, the
events are split into two jets using the Durham jet algorithm.
To ensure that the jet assignment is correct, each jet
must contain at least a charged track ($N_{\mathrm ch,j} \ge 1$), 
have a significant energy ($E_{\mathrm j}$)
and the magnitude of the sum of the track charges 
($|Q_{\mathrm j}|$) must not exceed 1.
If the acoplanarity angle is small,
cuts are applied on 
the transverse momentum ($P_{\mathrm t}$), the 
transverse momentum perpendicular to the event thrust axis ($a_{\mathrm t}$),
and $|\cos \theta_a|$ ($\theta_a \equiv \tan^{-1}(a_{\mathrm t}/P_z)$).
These cuts reduce the background from 
two-photon processes and lepton pairs.
If the acoplanarity angle is large, cuts on 
$P_{\mathrm t}$ and $| \cos\theta_{\mathrm miss}|$
are applied to reduce the two-photon background and reject events that
may have particles escaping detection along the beam line.   
The $P_{\mathrm t}$ distributions for the full acoplanarity angle region  
are shown in Fig.~\ref{catc1} after the cut on the jet charge.

\begin{figure} \centering
\begin{minipage}{16.0cm}
\begin{center}\mbox{
\epsfig{file=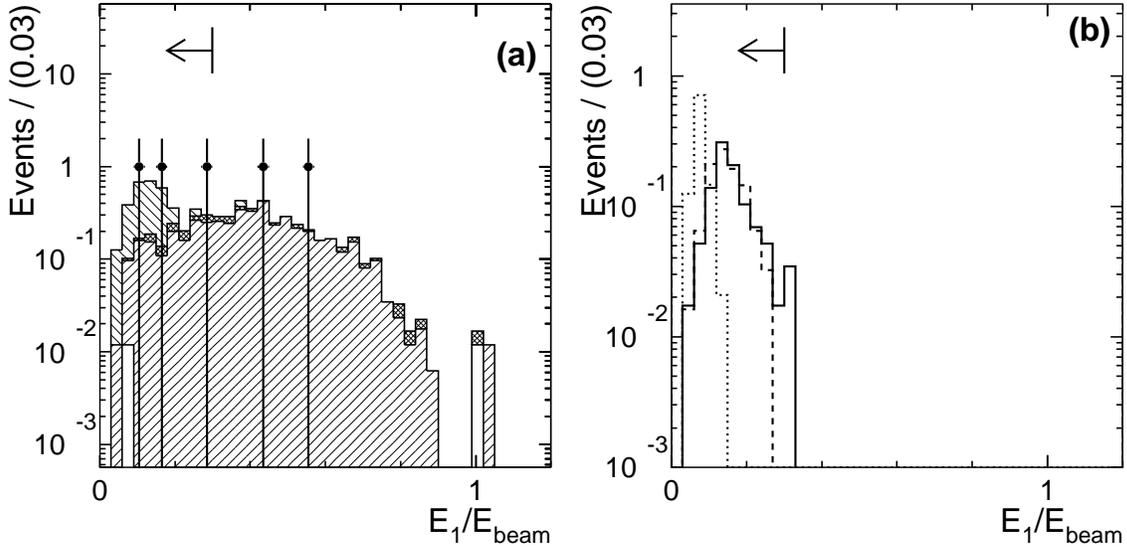,width=16.0cm}
}\end{center}
\vspace{-9mm}
\caption[]
{
Distributions of the highest jet energy in
analysis (C) region II after all other cuts have been applied.
Data and background contributions are shown in (a). The
background sources are shaded as in Fig.~\ref{fig:cataacop}.
The Monte Carlo signal distributions are shown in (b) and 
are labelled as in Fig.~\ref{catc1}.
The normalisations of the signal distributions are arbitrary. 
The arrows shown indicate the selection criteria.
}
\label{catc3}
\end{minipage}
\end{figure}

To reduce the background from $\ee\mumu$ events in which one of the muons 
is emitted at a  small polar angle and is not reconstructed as a good
track, events are rejected if there is a track segment in the
muon chamber, or a hadron calorimeter cluster at a small polar angle,
near the
missing momentum direction ($\vec P_{\mathrm miss}$) in the plane 
perpendicular to the beam axis. 
Soft hadronic events from two-photon processes are rejected by 
cuts on $|\cos \theta_{\mathrm j}|$, $\phi_{\mathrm acop}$ and  
$M_{\mathrm vis}$.
$\WW \ra \ell^+ \nu \ell^- \bar{\nu}$ events are rejected by
upper cuts on $M_{\mathrm vis}$ and
on the higher energy of the two jets, $E_1 / E_{\mathrm beam}$.
The distributions of $E_1 / E_{\mathrm beam}$  are shown for region II
in Fig.~\ref{catc3} after all the other cuts have been applied.
The values and ranges for all cuts are given in Table~\ref{tab:cutCch}.
The typical detection efficiencies for $\chp_1 \chm_1$ events 
are shown in Fig.~\ref{effi}a\@.

The analysis is especially optimised for the case that both charginos 
decay leptonically into three particles ($\chp_1 \ra \ell^+ \nu \nt_1$).
In addition the analysis is designed to have high efficiency in
the small $\Delta M_+$ region, where a large fraction of hadronic events 
fall into category (C).
If $m_{\snu}$ is smaller than $m_{\chp_1}$,
the two-body decays of the chargino ($\chp_1 \ra \snu \ell^+$)
may dominate over the three body decays via a virtual W.
The analysis of Ref. \cite{LEP18-slepton} 
especially tuned for 
acoplanar lepton search 
is applied in this case. 

\subsection{Detection of Neutralinos}

To obtain optimal performance the event sample is divided exclusively 
into two categories, motivated by the topologies expected from neutralino
events. 
\begin{itemize}
\item[(C)] $N_{\mathrm ch} \leq 4$:
the signal events in which $\nt_2$ decays into $\nt_1 \ellell$ tend
to fall into this category.
When the mass difference 
between $\nt_2$ and $\nt_1$ ($\Delta M_0 \equiv m_{\nt_2} - m_{\nt_1}$) is small,
signal events also tend to fall into this category.
\item[(D)] $N_{\mathrm ch} > 4$: the signal events
in which $\nt_2$ decays into $\nt_1 {\mathrm q}\bar{\mathrm q}$
tend to fall into this category for modest and large values of 
$\Delta M_0$\@.
\end{itemize}

For events with $N_{\mathrm ch} \le 4$ the category (C) cuts of the 
chargino search are used.
Events falling into category (D) have a monojet or di-jet topology and the
cuts described below provide better detection-performance for $\nt_1\nt_2$
detection than would have been obtained using the cuts of
category (A) of the chargino search.

The fraction of
events falling into
category (C) is 10-20\% for $\Delta M_0 \geq$20~GeV but
increases to about 70\% when  $\Delta M_0 \leq$5~GeV.
The fraction of invisible events due to $\nt_2 \ra \nt_1\Zrv \ra \nt_1 \nunu$ 
decays is 20-30\% depending on $\Delta M_0$\@.

The event shape of $\nt_1 \nt_2$ events mainly depends on the 
difference between the $\nt_2$ mass and the $\nt_1$
mass, therefore the selection criteria are optimised for four
$\Delta M_0$ regions:  
\begin{itemize}
\item[(i)]
$\Delta M_0 \leq 10$~GeV,
\item[(ii)]
$10 < \Delta M_0 \leq 30$~GeV,
\item[(iii)]
$30 < \Delta M_0 \leq 80$~GeV,
\item[(iv)]
$\Delta M_0 > 80$~GeV.
\end{itemize}
In regions i and ii, the main sources of background events are 
two-photon processes and the $\gamma^* Z^{(*)} \ra \qq \nunu$ processes.
In regions iii and iv, the main sources of background are
four-fermion processes ($\WW$, $\Wenu$ and $\gamma^* Z^{(*)}$).
Selection criteria applied for the low-multiplicity events (category (C))  
for regions i, ii, iii  and iv are identical to
those used in the analysis (C) of the chargino search
for regions Ia, II, III and IV, respectively (see Table~\ref{tab:cutCch}). 
The typical efficiencies for detecting $\nt_1 \nt_2$ events 
with the $\nt_2 \ra \nt_1 \Zv$ decay are shown in Fig.~\ref{effi}b\@.


\subsubsection{Analysis (D) ($\bf N_{\mathrm ch} > 4$)}

In $\nt_1 \nt_{2}$ events,
if the $\nt_{2}$ decays hadronically,
events tend to fall into category (D).
The variables used in the selection criteria and their cut values are 
listed in Table~\ref{tab:cutDnt}.

\begin{figure} \centering
\begin{minipage}{16.0cm}
\begin{center}\mbox{
\epsfig{file=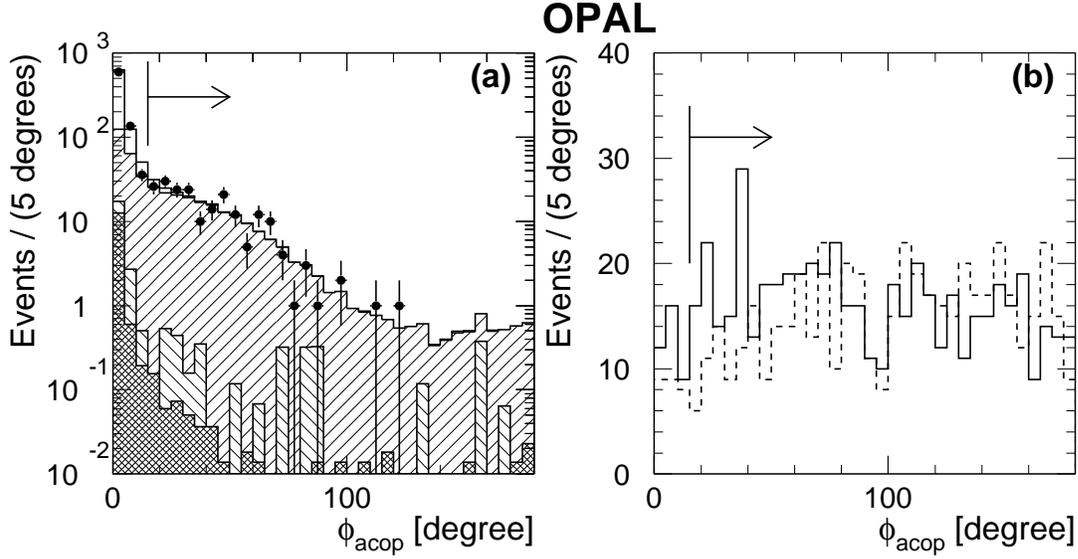,width=16.0cm}
}\end{center}
\vspace{-9mm}
\caption[]
{
The distributions of acoplanarity angle 
in analysis (D) region ii after cut on $P_t$\@. 
Data and background contributions are shown in (a)\@. 
The background sources are shaded as in Fig.~\ref{fig:cataacop}.
The Monte Carlo signal distributions are shown in (b) 
for $m_{\nt_2} = 105$~GeV and $m_{\nt_1} = 75$~GeV (solid line)
and for $m_{\nt_2} = 100$~GeV and $m_{\nt_1} = 80$~GeV (dashed line).
The normalisations of the signal distributions are arbitrary. 
The arrows shown indicate the selection criteria.
}
\label{catd1}
\end{minipage}
\end{figure}

To reduce the background from $\ee \ra \Zg$ 
and two-photon processes, requirements on $\thmiss$, 
$E_{\mathrm fwd}/E_{\mathrm vis}$ and transverse momenta are set.
The acoplanarity angle, as defined in the chargino analysis (A),
should be large to remove 
the two-photon and the $\qq$ background events.
To ensure the reliability of the measurement of $\phi_{\mathrm acop}$,
both jets should have a polar angle $\theta_{\mathrm j}$ in the range 
$|\cos\theta_{\mathrm j} |<0.95$.
The acoplanarity angle distribution for region ii 
after the $| \cos \theta_{\mathrm j}|$ cut is shown in Fig.~\ref{catd1}\@. 
For region iv the $\phi_{\mathrm acop}$ cut is loosened
with respect to regions i-iii, 
since the acoplanarity angle of signal events is smaller. 
On the other hand the $E_{\mathrm fwd}/E_{\mathrm vis}$ cut 
is tightened to reduce the $\qq$ background.

After these cuts, the remaining background events come 
predominantly from
$\gamma^* \Z \ra \qq \nunu$, $\WW \ra \ell \nu\qq^{'}$
and $\Wenu \ra  \qq ^{'}{\mathrm e} \nu$.
Cuts on the visible mass and on the ratio of the visible mass to the
visible energy are then applied to reduce the background from
 $\gamma^* Z \ra \qq \nunu$.
In regions iii and iv, 
$d^2_{23} \equiv y_{23} E_{\mathrm vis}^2 < 30$~GeV$^2$ is required  
to select a clear two-jet topology and to reject 
$\WW \ra \tau \nu \qq'$ events.
In Fig.8 the $d^2_{23}$ distribution is shown for region iv
after all the other cuts. 
Typical detection efficiencies for $\nt_2 \nt_1$ events are shown 
in Fig.~\ref{effi}b\@.

\begin{table}[htb]
\centering
\begin{tabular}{|l||r|r|r|r|}
\hline
Region  & i~~ & ii~ & iii & iv \\
\hline
\hline
$E_{\mathrm fwd}/E_{\mathrm vis}$ 
& \multicolumn{2}{|c|}{$<0.15$} & $<0.20$ & $<0.05$   \\ 
\hline
$|\cos \theta_{\mathrm miss}|$ 
& \multicolumn{2}{|c|}{$<0.8$} & \multicolumn{2}{|c|}{$<0.9$}  \\ 
\hline
$p_{\mathrm t}$~GeV & \multicolumn{2}{|c|}{$>5$} 
                    & \multicolumn{2}{|c|}{$>7$}  \\  
$p_{\mathrm t}^{\mathrm HCAL}$~GeV & \multicolumn{2}{|c|}{$>5$}
                                   & \multicolumn{2}{|c|}{$>7$} \\  
\hline
$|\cos \theta_{\mathrm j}|$ 
& \multicolumn{4}{|c|}{$<0.95$} \\ 
\hline
$\phi_{\mathrm acop}^{\circ}$ & \multicolumn{3}{|c|}{$>20$} & $>10$   \\
\hline
$M_{\mathrm vis}$~GeV & $<12$ & $<35$& $<70$ & [20,130]  \\  
\hline
$M_{\mathrm vis}/E_{\mathrm vis}$ &  \multicolumn{3}{|c|}{$>0.3$} & --   \\  
\hline
$d^2_{23}$~GeV$^2$ & \multicolumn{2}{|c|}{--} 
                   & \multicolumn{2}{|c|}{$<30$} \\  
\hline
\hline
background &   &   &  &     \\  
\hline
$\gamma \gamma$     & 1.05 & 2.15 & 0.77 & 0.00 \\ 
\hline
$\ellell (\gamma)$  & 0.00 & 0.03 & 0.07 & 0.10  \\  
\hline
$\qq (\gamma)$      & 0.01 & 0.01 & 0.04 & 0.07  \\  
\hline
4f                  & 0.12 & 1.29 & 3.18 & 10.37 \\  
\hline
\hline
total bkg & 1.19$\pm$0.50 & 3.49$\pm$0.77 & 4.05$\pm$0.51 & 10.55$\pm$0.35 \\  
\hline
\hline
observed       & 0  &  1  &  3 & 12   \\  
\hline
\end{tabular}
\caption[]{
  \protect{\parbox[t]{12cm}{
The list of selection criteria for $\nt_2 \nt_1$ events in category (D).
The numbers of background events expected for the integrated luminosity 
of $\lumi$~pb$^{-1}$ for various Standard Model processes and the total
number of background events expected as well as the observed number of 
events in each $\Delta M_0$ region are also listed.
The errors in the total background include only the statistical error
of simulated background events.
}} 
}
\label{tab:cutDnt}
\end{table}
 
\subsection{Systematic errors and corrections }
Systematic errors on the number of expected signal events arise from 
the following sources:
the measurement of the integrated luminosity (0.5\%),
Monte Carlo statistics in the various signal samples, 
interpolation errors of the efficiencies to arbitrary values
of $m_{\chp_1}$ and $m_{\nt_1}$ (2--10\%),
modelling of the cut variables in the Monte Carlo
simulations\footnote{
This is estimated by comparing the efficiencies obtained by shifting
each cut variable by the possible  shift in the corresponding
distribution which still gives agreement between data and Monte
Carlo.} 
(4--10\%), errors
due to fragmentation uncertainties in hadronic decays ($< 2$\%),
the matrix elements leading to 
different decay parameters ($< 5$\%)
and effects of detector calibration ($< 1$\%).
The effect of possible trigger inefficiencies
has been investigated and found to be negligible.

Systematic errors on the expected number of background
events are due to
Monte Carlo statistics in the simulated background events (as quoted
in Tables~\ref{tab:cutAch}, \ref{tab:cutBch}, \ref{tab:cutCch} and 
\ref{tab:cutDnt}),
uncertainties in the amount of two-photon background which are estimated by
fitting the $P_t$ distributions of simulated two-photon events
and the data (30\%), and uncertainties in the simulation of
the four-fermion processes which are estimated by taking the difference
between the predictions of the grc4f~\cite{grc4f} and the 
EXCALIBUR~\cite{excalibur} generators (17\%). 
The systematic errors due to
the modelling of the cut variables in the detector simulation are 
less than 7\%.


The rate of events in which the measured energy  in the
SW, FD or GC calorimeters, 
due to noise and beam related background, 
exceeded the thresholds in the preselection
is 4.5\%  as estimated
from random beam crossing events. Since this effect is not modelled in
the simulation, it is taken into account by scaling down 
the integrated luminosity by this amount.

\section{Results}
\subsection{Limits on the $\chp_1 \chm_1$ 
and $\nt_2 \nt_1$ production cross-sections}

Model-independent upper limits are obtained at 
95\% C.L. on the production cross-sections.  
%
This is done for $\chp_1 \chm_1$ assuming
the specific decay mode $\chpm_1 \ra \nt_1 {\mathrm W}^{(*)\pm}$ 
and for $\nt_1 \nt_2$ production assuming the 
$\nt_2 \ra \nt_1 {\mathrm Z}^{(*)}$ decay.
Exclusion regions are determined from
the observed numbers of events 
at $\roots =$181--184~GeV\footnote{
When calculating limits, cross-sections at different $\roots$ are estimated
by weighting by $\bar\beta / s$ 
in these proportions, where $\bar\beta$ is 
$p_{\chpm_1}/E_{\mathrm beam}$ for $\chp_1 \chm_1$ production 
or  $p_{\nt_2}/E_{\mathrm beam} = p_{\nt_1}/E_{\mathrm beam}$ for 
$\nt_2 \nt_1$ production.},
the signal detection efficiencies and their uncertainties,
and the numbers of background events expected and their uncertainties.
To obtain this limit at a given 
($m_{\chp_1}$, $m_{\nt_1}$) or ($m_{\nt_2}$, $m_{\nt_1}$) point,
the independent analyses ((A), (B) and (C) for chargino, (C) and (D) 
for neutralino) are combined using the likelihood ratio method~\cite{LH}.
This method assigns 
greater weight to the analysis which has greater sensitivity.  

Systematic uncertainties on the efficiencies are incorporated
following the method in Ref.~\cite{Cousins}, and the
systematic uncertainties on
the number of expected background events are
incorporated by numerical integration, assuming Gaussian errors,
as suggested in Ref.~\cite{Cousins}. 

Contours of the 95\% C.L. upper limits for 
the $\chp_1 \chm_1$ cross-sections 
are shown in Fig.~\ref{figsum1} 
assuming the $\chpm_1 \ra \nt_1 {\mathrm W}^{(*)\pm}$ decay 
with 100\% branching fraction.
Although these limits do not depend on the details of the SUSY models
considered, a ``typical" field
content of the gauginos is assumed, leading to particular 
production angular distributions that are subsequently used
in estimating detection efficiencies.
Differences in detection efficiencies arise from
variations in the angular distributions 
obtained by using different MSSM
parameters corresponding to the same mass combination.
The variation of the efficiency is observed to be less than 2\%.
Of the parameters examined, those yielding
the lowest efficiency are used.
If the cross-section for $\chp_1 \chm_1$ is larger than
0.6~pb and $\Delta M_+$ is between 5~GeV and about 80~GeV,
$m_{\chp_1}$ is excluded with 95\% C.L. up to the kinematic limit, 
assuming $Br(\chp_1 \ra \nt_1 {\mathrm W}^{(*)+}) = 100\%$\@.

Similar contours of the upper limits for
the $\nt_2 \nt_1$ cross-sections 
are shown in Fig.~\ref{figsum2}.
If the cross-section for $\nt_2 \nt_1$ is larger than
0.3~pb and $\Delta M_0$ is greater than 10~GeV, 
$m_{\nt_2}$ is excluded up to
the kinematic limit at 95\% C.L.,
assuming $Br(\nt_2 \ra \nt_1 {\mathrm Z}^{(*)}) = 100\%$\@.

\subsection{Limits in the MSSM parameter space}
\label{interp}
The results of the above searches can be interpreted within the
framework of the Constrained MSSM (CMSSM).
The phenomenology of the gaugino-higgsino
sector of the MSSM is mostly determined by the 
parameters $M_2$, $\mu$ and $\tan\beta$.
In the absence of light sfermions and light SUSY Higgs
particles, these three parameters are sufficient to describe
the chargino and neutralino sectors completely.  
Within the CMSSM, 
a large value of the common 
scalar mass, $m_0$ (e.g., $m_0 = 500$~GeV) leads to heavy sfermions
and therefore to a negligible 
suppression of the cross-section due to interference 
from $t$-channel sneutrino exchange. Chargino 
decays would then proceed predominantly
via a virtual or real W.
On the other hand, a light $m_0$ results in a low value of
the mass of the $\snu$,
enhancing the contribution of the $t$-channel exchange diagrams 
that may have destructive interference with $s$-channel diagrams,
thus reducing the cross-section for chargino pair production.
Small values of $m_0$ also tend to enhance the leptonic branching
ratio of charginos, often leading to smaller detection efficiencies.
Certain values of $m_0$ can lead to the condition $m_{\snu} < m_{\chpm_1}$
and result in the two-body decay mode $\chpm_1 \ra \snu \ell^{\pm}$
being dominant.   The chargino detection efficiency can be small
or zero for these decays, 
particularly when $m_{\snu} \approx m_{\chpm_1}$, 
leading to severe degradation in sensitivity.


From the input parameters $M_2$, $\mu$, $\tan \beta$,
$m_0$ and $A$ (the trilinear coupling),
masses, production cross-sections and
branching fractions are calculated
according to the CMSSM~\cite{Bartl,chargino-theory,Ambrosanio,Carena}. 
For each set of input parameters, the total number of
$\chp_1 \chm_1$, $\nt_2 \nt_1$, $\nt_3 \nt_1$ and $\nt_2 \nt_2$ events
expected to be observed are calculated using
the integrated luminosity, the cross-sections,
branching fractions, and 
the detection efficiencies which depend upon the masses of the chargino,
the lightest neutralino and next-to-lightest neutralino.
The relative importance of each of the analyses (A)--(D) changes with
the leptonic or hadronic branching ratios, and the likelihood
ratio method~\cite{LH} is used to optimally weight each analysis
depending on these branching ratios.

Results are presented for two cases: (i) $m_0 = 500$~GeV (i.e., heavy 
sfermions), 
and (ii) the value of $m_0$
that gives the smallest total number of expected chargino
and neutralino events 
taking into account cross-sections,
branching ratios, and detection efficiencies
for each set of
values of $M_2$, $\mu$, $\tan \beta$.
This value of $m_0$ hence leads to the worst limit at that point,
so that the resulting limits are valid for all $m_0$.
Values of $m_0$ are considered that 
remain compatible with the current limits
on the $\snu$ mass ($m_{\snu_L} > 43$~GeV~\cite{PDG98}), and
OPAL upper limits on the cross-section for $\tilde{\ell}$ pair production, 
particularly
right-handed smuon and selectron pair 
production~\cite{LEP18-slepton}.
Particular attention is paid to the region of values of $m_0$ 
leading
to the critical mass condition
$m_{\snu} \approx m_{\chpm_1}$.
When $m_{\snu} \leq m_{\chpm_1}$, resulting in a topology of 
acoplanar leptons and missing momentum, the upper limits
on the cross-section for the two-body chargino decay from
Ref.~\cite{LEP18-slepton} are used.
The contribution of the cascade decays $\chp_1 \ra \nt_2 \mathrm{X}$ followed by
$\nt_2 \ra \nt_1 \mathrm{Y}$ are also included.
The photonic radiative decay $\nt_2 \ra \nt_1 \gamma$ leading to single
photon topologies from $\nt_2 \nt_1$ production and acoplanar photons
with missing energy topologies from $\nt_2 \nt_2$ are taken into account
using the 95\% C.L. cross-section upper limits on these topologies from
the OPAL results of Ref.~\cite{gamgam}.
In both of these cases, if the relevant product of
cross-section and branching ratio for a particular set
of MSSM parameters is greater than the measured 95\% C.L. upper limit
presented in that paper, then that set of parameters is considered to be
excluded.

The following regions of the CMSSM parameters are scanned:
$0\le M_2 \le 2000$~GeV,
$|\mu| \le 500$~GeV, and
$A = \pm M_2, \,\, \pm m_0$ and 0.
The typical scan step is 0.2~GeV.
Checks have been made to ensure that
the scanned ranges of parameters are
large enough that the exclusion regions change
negligibly for even larger ranges.
No significant dependence on $A$ is observed.
The 95\% C.L. upper limit on the expected number of 
events is determined and systematic errors on efficiencies and
backgrounds are incorporated as described previously.
Figure~\ref{fig_mssm} 
shows the resulting exclusion regions 
in the ($M_2$,$\mu$) plane
for $\tan\beta = 1.5$ and 35 with $m_0 \geq 500$~GeV and
for all $m_0$.

The restrictions on the CMSSM parameter space presented here can be
transformed into exclusion regions in the ($m_{\chpm_1}$,$m_{\nt_1}$)
or ($m_{\nt_2}$,$m_{\nt_1}$) plane.
A given mass pair is excluded only if
{\it all} CMSSM parameters in the scan which lead
to that same mass pair are
excluded at the 95\%~C.L.
The $\chpm_1$ mass limits are summarised 
in Table~\ref{tab:resultsc}. 
In the ($m_{\chpm_1}$,$m_{\nt_1}$) plane, Fig.~\ref{masslimc}
shows the corresponding
95\% C.L. exclusion regions for $\tan\beta = 1.5$ and 35.

Figure~\ref{masslimn} shows the corresponding
95\% C.L. exclusion regions in the ($m_{\nt_2}$,$m_{\nt_1}$) 
plane, for $\tan\beta = 1.5$ and 35.
Although the neutralino production cross-section is small, our
detection efficiencies in the direct neutralino search are high
enough, and we have now collected enough integrated luminosity that,
for certain SUSY parameters, we should be able to observe them
directly rather than their being excluded indirectly through the
exclusion of charginos of certain masses.
Regions that would be excluded by the direct neutralino searches
alone are also shown in Fig.~\ref{masslimn} delimited by the white
dotted lines.
Mass limits on $\nt_1$, $\nt_2$, and $\nt_3$ 
are summarised in Table~\ref{tab:resultsn}.

\begin{table}[htb]
\centering
\begin{tabular}{|c|c||c|c|}
\hline
 &     & $\tan \beta = 1.5$  & $\tan \beta = 35$ \\
\hline
\hline
$m_0 \ge 500$~GeV &
$\Delta M_+ \geq 5$~GeV     & $m_{\chp_1}>90.0$~GeV & $m_{\chp_1}>90.2$~GeV \\

   &
$\Delta M_+ \geq 10$~GeV    & $m_{\chp_1}>91.1$~GeV & $m_{\chp_1}>91.2$~GeV \\
\hline
All $m_0$ (see text) &
$\Delta M_+ \geq 5$~GeV    & $m_{\chp_1}>69.1$~GeV & $m_{\chp_1}>65.2$~GeV \\
\hline
\end{tabular}
\caption[]{
   \protect{\parbox[t]{12cm}{
Lower limits at 95\% C.L. obtained on the lightest chargino mass.

}} }
\label{tab:resultsc}
\end{table}

\begin{table}[htb]
\centering
\begin{tabular}{|c|c||c|c|}
\hline
   &   & $\tan \beta = 1.5$  & $\tan \beta = 35$ \\
\hline
\hline
$m_0 \ge 500$~GeV & No $\Delta M_0$ restriction& 
       $m_{\nt_1}>38.1$~GeV & $m_{\nt_1}>46.4$~GeV \\ \cline{2-4}
 & $\Delta M_0 \geq 10$~GeV   & $m_{\nt_2}>63.0$~GeV & $m_{\nt_2}>91.1$~GeV \\
 &                            & $m_{\nt_3}>102.3$~GeV  & $m_{\nt_3}>122.1$~GeV \\
\hline
All $m_0$ (see text) & No $\Delta M_0$ restriction& 
       $m_{\nt_1}>25.4$~GeV & $m_{\nt_1}>36.5$~GeV \\ \cline{2-4}
 & $\Delta M_0 \geq 10$~GeV   & $m_{\nt_2}>51.7$~GeV & $m_{\nt_2}>84.8$~GeV \\
 &                            & $m_{\nt_3}>102.1$~GeV  & $m_{\nt_3}>124.4$~GeV \\
\hline
\end{tabular}
\caption[]{
   \protect{\parbox[t]{12cm}{
Lower limits at 95\% C.L. obtained on $m_{\nt_1}$, $m_{\nt_2}$, and
$m_{\nt_3}$. 
}} }
\label{tab:resultsn}
\end{table}



To study the sensitivity of these numerical mass limits, the
limits expected from the Standard Model background processes
are computed assuming no signal.
For chargino mass limits close to the kinematic limit,
the expected limits differ by less
than 0.2~GeV from the observed limit values, while
expected mass limits for other values
in Table~\ref{tab:resultsc} and Table~\ref{tab:resultsn} differ
by less than $3.0$~GeV from the observed values in all cases.

In the ``Higgsino region'' where $M_2$ is large, the mass difference
$\Delta M_+$ between the chargino and lightest neutralino decreases
with increasing $M_2$, resulting in a drop in detection efficiency.
For $\tan\beta = 1.5$,
Fig.~\ref{plotm2} illustrates the gaugino mass limits in 
slices of constant $M_2$ as well as the correspondence of
$M_2$ with mass difference $\Delta M_+$.  For $\mu < 0$
the $m_{\chp_1}$ limit is above the kinematical boundary of  
$\chp_1 \chm_1$ production.
This is obtained from the interpretation of the results of the direct 
$\nt_2 \nt_1$ searches.
For larger values of $\tan\beta$ the form of the curves are similar, but
the difference between the limits for $\mu < 0$ and $\mu > 0$ for
a particular gaugino decreases, and  the limits are in general 
between the $\mu < 0$ and $\mu > 0$ curves shown for $\tan\beta = 1.5$. 

Figure~\ref{limchsnu} illustrates the sensitivity of the chargino
mass limit
to the mass of the sneutrino set by the chosen value of $m_0$.
The dashed line delineates the condition $m_{\chpm_1} = m_{\tilde{\nu}}$
where the $\chp_1 \chm_1$ search fails due to small visible energy. 
In the region to the right side of the line, charginos will undergo
three-body decays
which are searched for directly, while on the left side of the
line they
decay via the two-body mode $\chpm \ra \tilde{\nu} \ell^{\pm}$
and the results of Ref.~\cite{LEP18-slepton} are applied.
In this case the chargino lower mass limit is set mostly by the
right-handed slepton limits with some contributions from neutralino
production.

Figure~\ref{limtanb} shows the dependence of the mass limits on 
the value of $\tan\beta$.  Of particular interest is the absolute
lower limit, in the framework of the CMSSM, on the mass of the lightest 
neutralino of $m_{\nt_1} > 30.1$~GeV ($24.2$~GeV) at 95\% C.L. 
for $m_0 \ge 500$~GeV (all $m_0$).
This has implications on direct searches 
for the lightest neutralino as a candidate for dark matter.
If the lightest neutralino forms dark matter and it is 
lighter than about 20~GeV, 
it is difficult to detect it by terrestrial dark matter searches 
due to the small recoil energy in the material~\cite{dark_matter}\@.
Since the formulae for couplings and masses in the gaugino sector
are symmetric in $\tan\beta$ and $1/\tan\beta$, these results
also hold for $\tan\beta < 1$.

\section{Summary and Conclusion}

A data sample corresponding to
an integrated luminosity of $\lumi$~pb$^{-1}$
at $\roots = $181--184~GeV,
collected with the OPAL detector, has been analysed
to search for pair production of charginos and neutralinos  
predicted by supersymmetric theories.
No evidence for these events has been observed.
Assuming $m_{\chpm_1}-m_{\nt_1} \geq 5$~GeV,
the 95\% C.L. lower mass limit of the chargino is 90.0~GeV
for $\tan \beta = 1.5$ and 90.2~GeV for $\tan \beta = 35$, 
within the framework of the CMSSM and  
for the case of a large universal scalar mass ($m_0 \geq$~500~GeV).
For all $m_0$, the 95\% C.L. lower mass limit 
is 69.1~GeV for $\tan \beta = 1.5$
and 65.2~GeV for $\tan \beta = 35$\@.
In certain regions of parameter space,
the $m_{\chpm_1}$ limit exceeds the kinematical 
boundary of the $\chp_1 \chm_1$ production due to the  
interpretation of the results from the direct $\nt_2 \nt_1$ search.
The absolute lower mass limit on the lightest neutralino
in the framework of the CMSSM with
$m_0 \ge 500$~GeV is 30.1~GeV at 95\% C.L., and
$24.2$~GeV for the worst-case limit with all $m_0$.
This has implications for experimental searches for 
the lightest neutralino as a dark matter candidate.

\section*{Acknowledgements}

We particularly wish to thank the SL Division for the efficient operation
of the LEP accelerator at all energies
and for their continuing close cooperation with
our experimental group.  We thank our colleagues from CEA, DAPNIA/SPP,
CE-Saclay for their efforts over the years on the time-of-flight and trigger
systems which we continue to use.  In addition to the support staff at our own
institutions we are pleased to acknowledge the  \\
Department of Energy, USA, \\
National Science Foundation, USA, \\
Particle Physics and Astronomy Research Council, UK, \\
Natural Sciences and Engineering Research Council, Canada, \\
Israel Science Foundation, administered by the Israel
Academy of Science and Humanities, \\
Minerva Gesellschaft, \\
Benoziyo Center for High Energy Physics,\\
Japanese Ministry of Education, Science and Culture (the
Monbusho) and a grant under the Monbusho International
Science Research Program,\\
Japanese Society for the Promotion of Science (JSPS),\\
German Israeli Bi-national Science Foundation (GIF), \\
Bundesministerium f\"ur Bildung, Wissenschaft,
Forschung und Technologie, Germany, \\
National Research Council of Canada, \\
Research Corporation, USA,\\
Hungarian Foundation for Scientific Research, OTKA T-016660, 
T023793 and OTKA F-023259.\\



\bibliographystyle{unsrt}


\newpage

\begin{figure} \centering
\begin{minipage}{16.0cm}
\begin{center}\mbox{
\epsfig{file=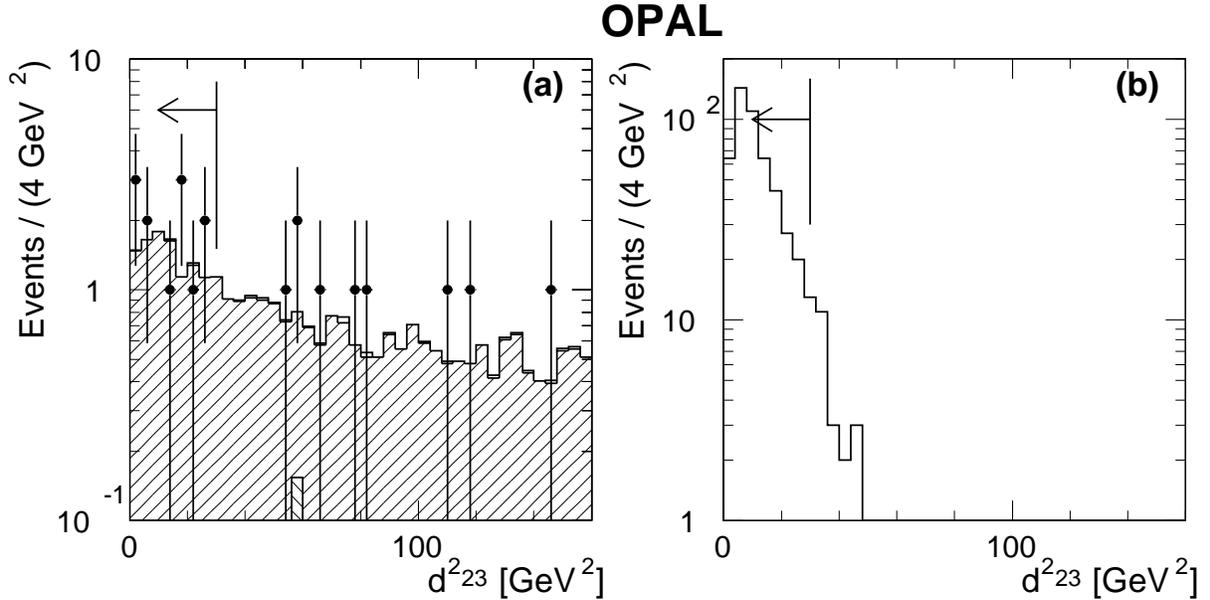,width=16.0cm}
}\end{center}
\vspace{-9mm}
\caption[]
{
The distributions of $d_{23} (\equiv y_{23} E_{\mathrm vis}^2)$~GeV$^2$
in analysis (D) region iv for three jet events after all the other cuts.
Data and background contributions are shown in (a). The
background sources are shaded as in Fig.~\ref{fig:cataacop}.
The Monte Carlo signal distribution is shown in (b) 
for $m_{\nt_2} = 150$~GeV and $m_{\nt_1} = 20$~GeV\@.
The normalisation of the signal distribution is arbitrary. 
The arrows shown indicate the selection criteria.
}
\label{catd2}
\end{minipage}
\end{figure}

\newpage

\begin{figure} \centering
\begin{minipage}{16.0cm}
\begin{center}\mbox{
\epsfig{file=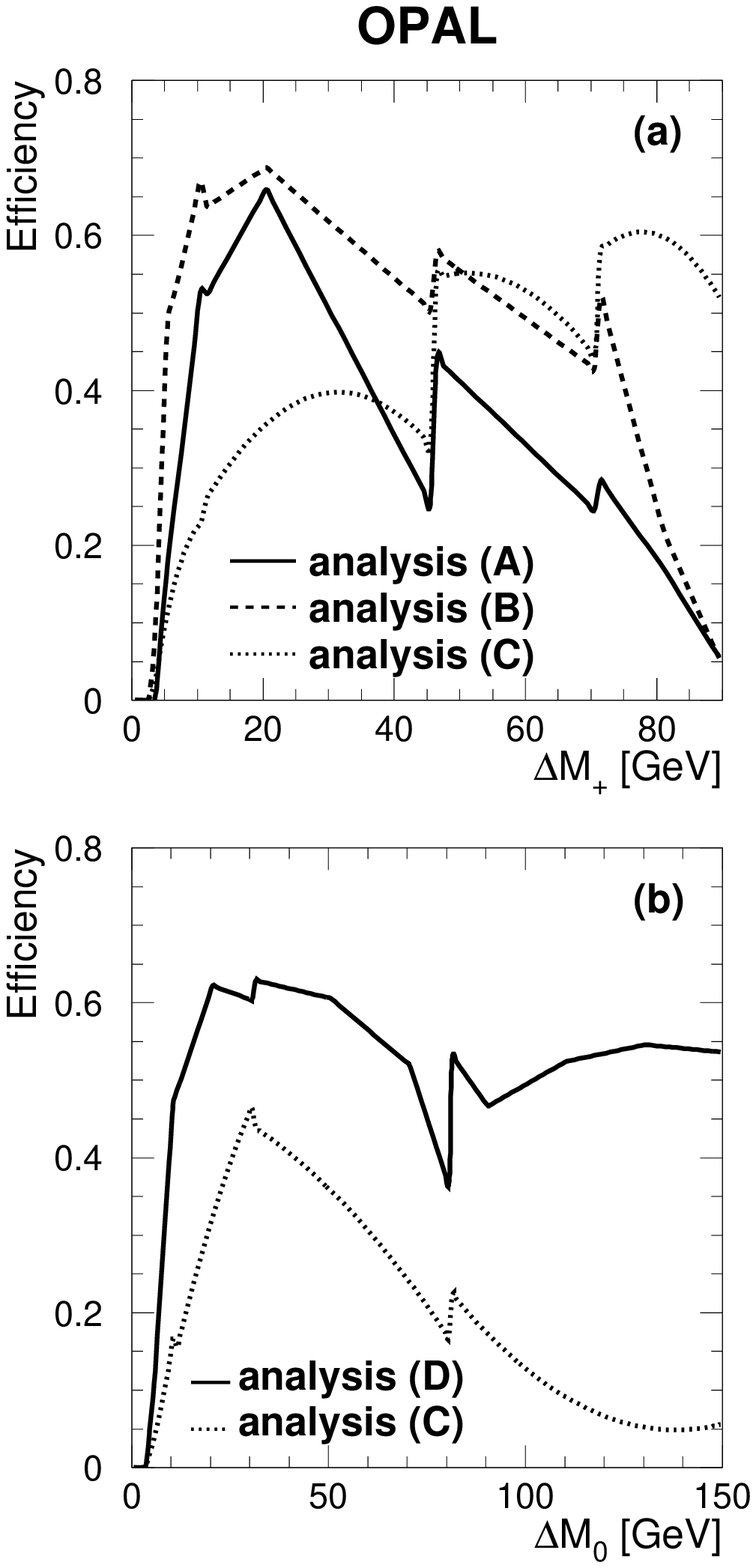,width=10.0cm}
}\end{center}
\vspace{-9mm}
\caption[]
{
(a) Detection efficiencies of analyses (A) (solid line), (B) (dashed line)
and (C) (dotted line) for $\chp_1 \chm_1$ events 
as a function of 
$\Delta M_+ \equiv m_{\chp_1} - m_{\nt_1}$, for
$m_{\chp_1} = 90$~GeV.
(b) Detection efficiencies of analyses (C) (dotted line) and (D) (solid
line) for
$\nt_1 \nt_2$ events as a function of
$\Delta M_0 \equiv m_{\nt_2} - m_{\nt_1}$, for
$m_{\nt_1} + m_{\nt_2} = 170$~GeV.
The efficiencies for each analysis are normalised to 
the number of events in each category.
}
\label{effi}
\end{minipage}
\end{figure}

\begin{figure} \centering
\begin{minipage}{16.0cm}
\begin{center}\mbox{
\epsfig{file=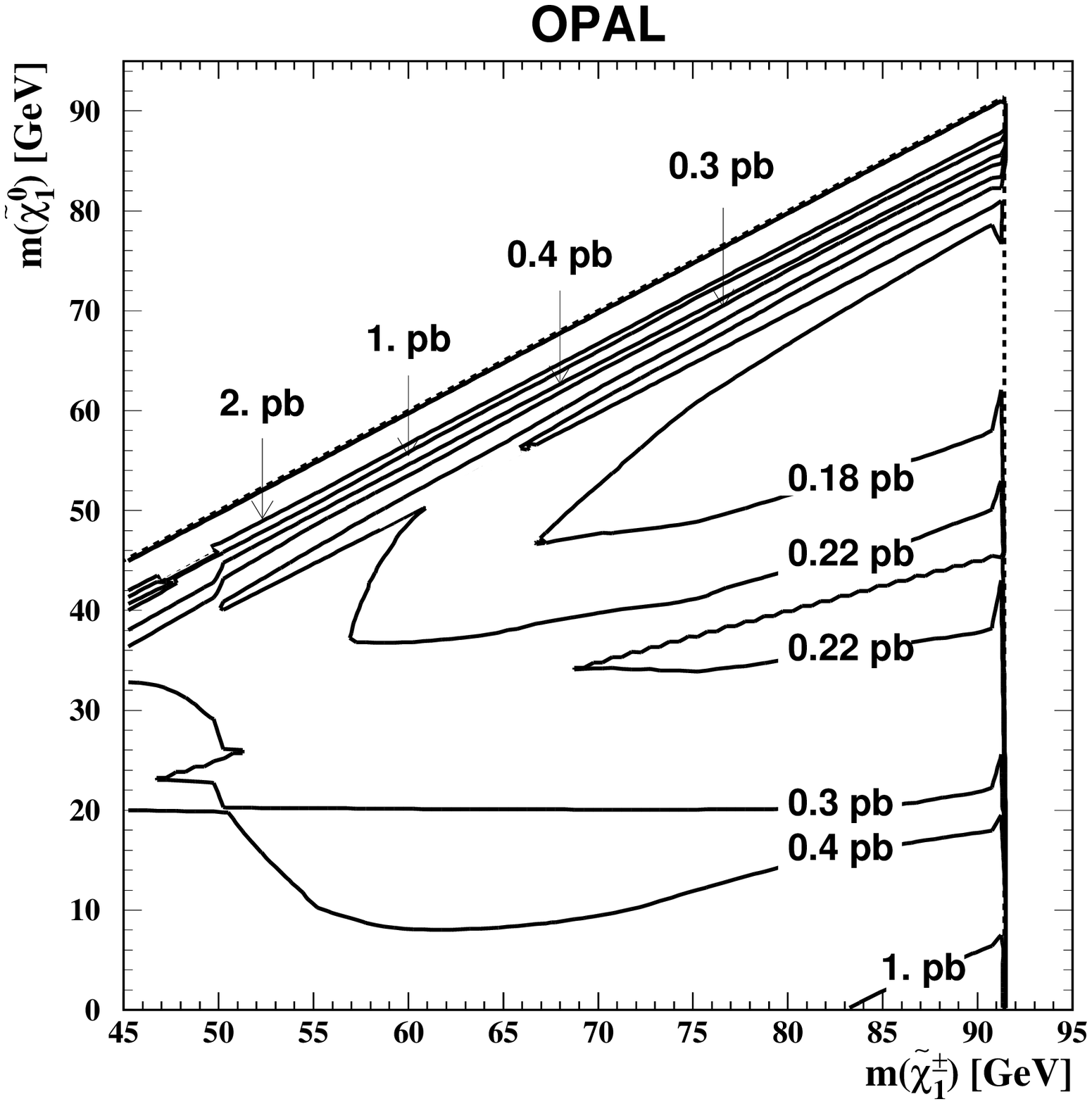,width=16cm,
bbllx=0pt,bblly=140pt,bburx=567pt,bbury=667pt}
}\end{center}
\vspace{-9mm}
\caption[]
{
The contours of the 95\% C.L. upper limits
for the $\ee \ra \chp_1 \chm_1$ production cross-sections at
$\roots =$ 182.7~GeV are shown
assuming $Br(\chp_1 \ra \nt_1 {\mathrm W}^{(*)+}) = 100$\%. 
}
\label{figsum1}
\end{minipage}
\end{figure}

\begin{figure} \centering
\begin{minipage}{16.0cm}
\begin{center}\mbox{
\epsfig{file=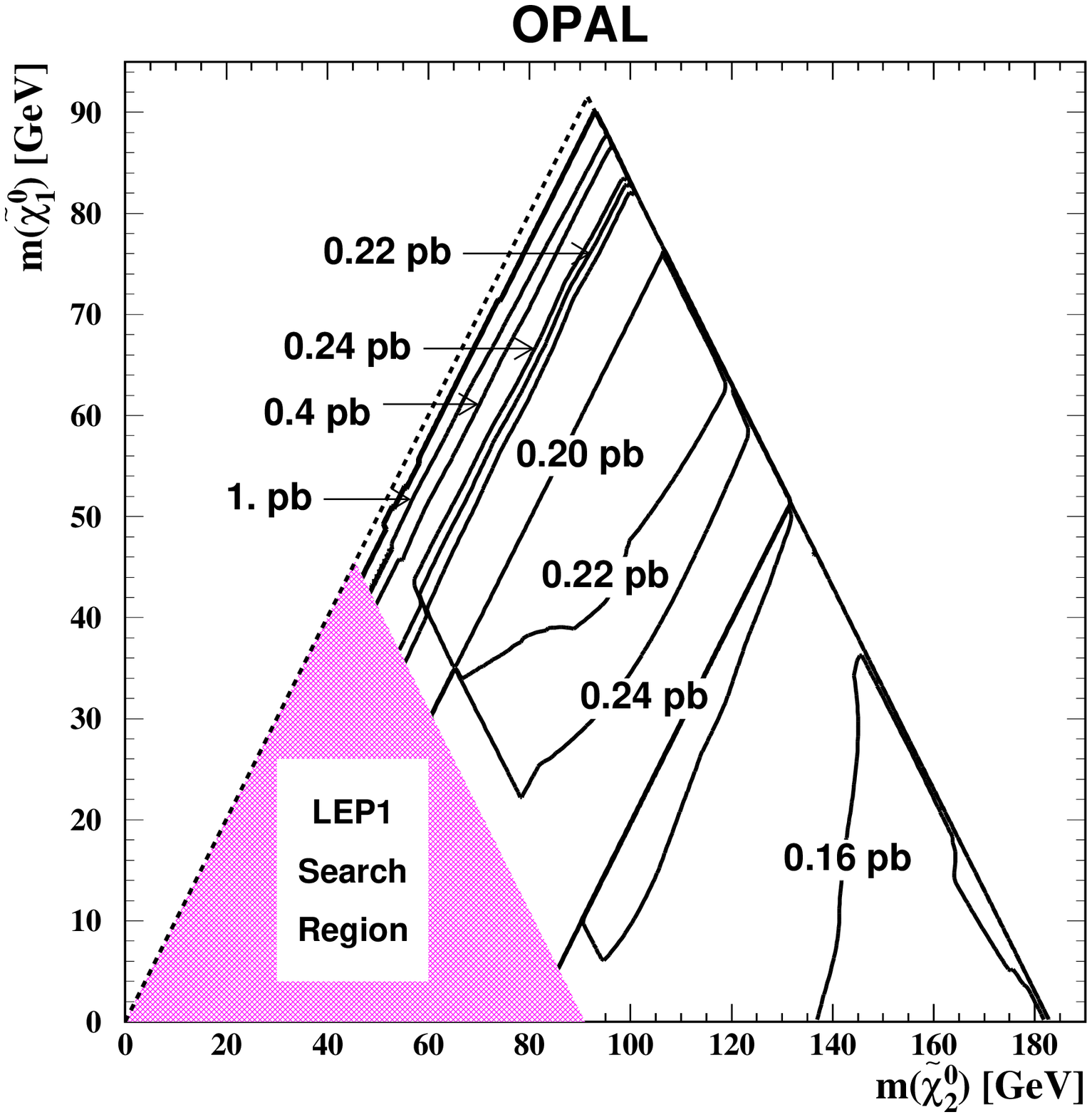,width=16cm,
bbllx=0pt,bblly=140pt,bburx=567pt,bbury=667pt}
}\end{center}
\vspace{-9mm}
\caption[]
{
The contours of the 95\% C.L. upper limits
for the $\ee \ra \nt_2 \nt_1$ production cross-sections at
$\roots =$ 182.7~GeV are shown
assuming $Br(\nt_2 \ra \nt_1 {\mathrm Z}^{(*)}) = 100$\%. 
The region for which $m_{\nt_2} + m_{\nt_1} < m_{\mathrm Z}$
is not considered in this analysis.
}
\label{figsum2}
\end{minipage}
\end{figure}

\begin{figure} \centering
\begin{minipage}{16.0cm}
\begin{center}\mbox{
\epsfig{file=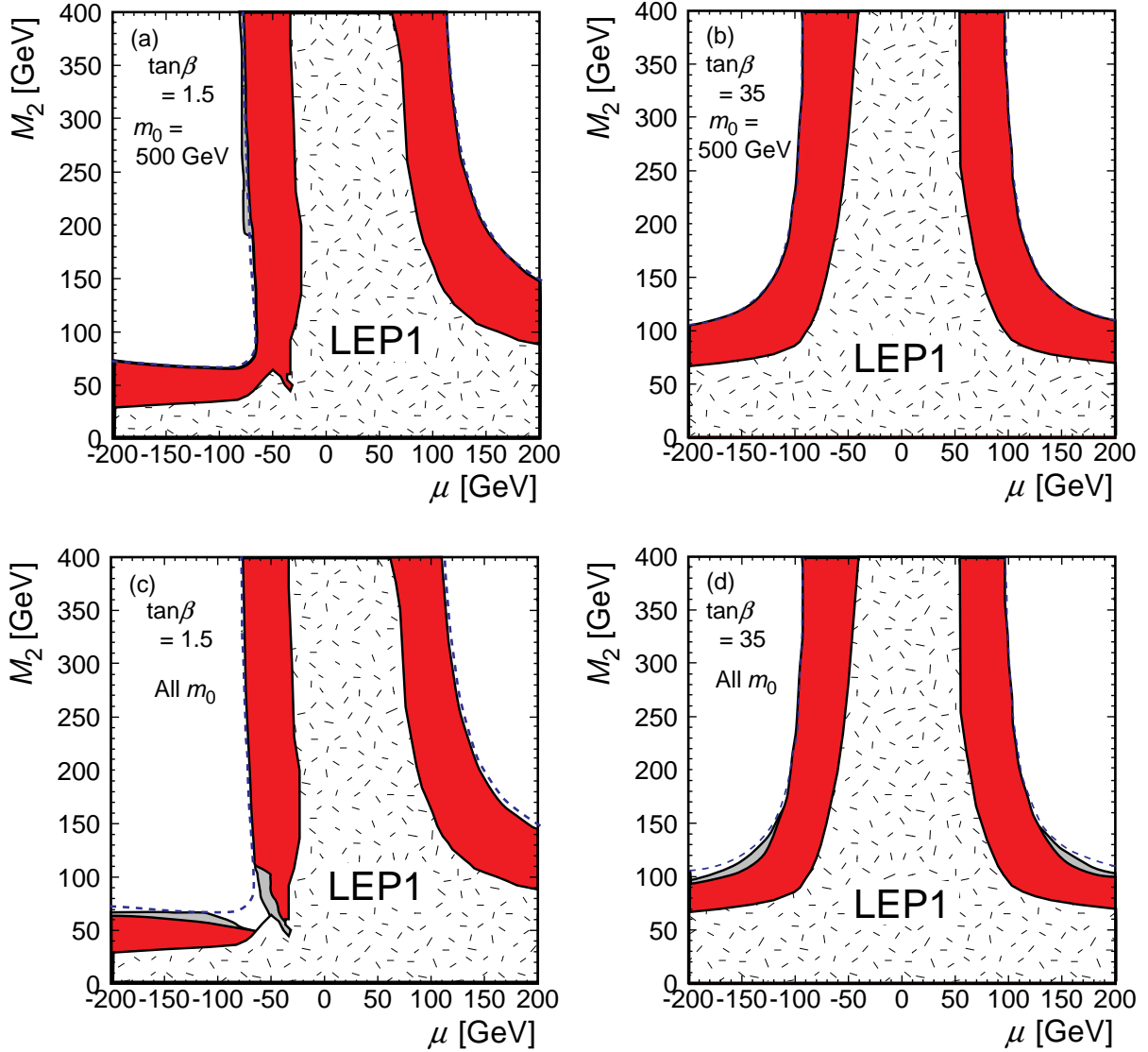,width=16.0cm}
}\end{center}
\vspace{-9mm}
\caption[]
{
Exclusion regions at 95\% C.L. in the ($M_2$,$\mu$) plane 
with $m_0 \geq 500$~GeV
for (a) $\tan \beta=1.5$ and for (b) $\tan \beta=35$.
Exclusion regions valid for all $m_0$  
for (c) $\tan \beta=1.5$ and for (d) $\tan \beta=35$.
The speckled areas show the LEP1 excluded regions
and the dark shaded areas show the additional exclusion
region using the data from $\sqrt{s}=181$--184~GeV.
The kinematical boundaries for $\chp_1 \chm_1$ production are shown
by the dashed curves.
The light shaded region in (a) extending beyond the  kinematical 
boundary of the $\chp_1 \chm_1$ production is obtained due to the  
interpretation of the results from the direct neutralino searches.
The light shaded regions
elsewhere show the additional exclusion regions due to
direct neutralino searches and other OPAL search results (see text).
}

\label{fig_mssm}
\end{minipage}
\end{figure}

\begin{figure} \centering
\begin{minipage}{16.0cm}
\begin{center}\mbox{
\epsfig{file=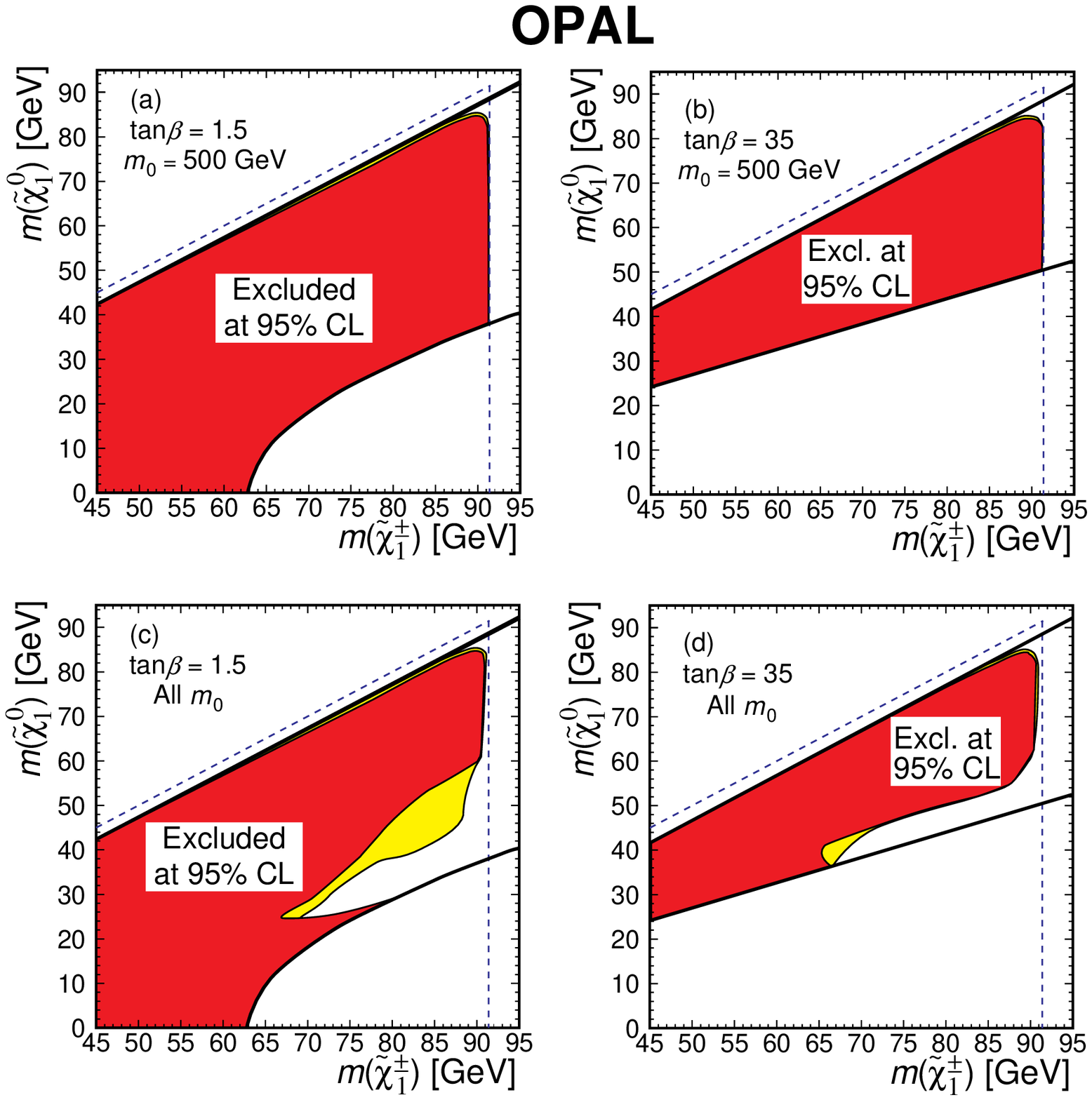,width=16.0cm}
}\end{center}
\vspace{-9mm}
\caption[]
{
The 95\% C.L. excluded region
in the ($m_{\nt_1}$,$m_{\chp_1}$) plane
within the framework of the CMSSM 
with $m_0 \geq 500$~GeV
for (a) $\tan \beta=1.5$ and for (b) $\tan \beta=35$.
The excluded region valid for all $m_0$  
for (c) $\tan \beta=1.5$ and for (d) $\tan \beta=35$.
The thick solid lines represent the theoretical bounds
of the CMSSM parameter space as given in the text.
The kinematical boundaries for $\chp_1 \chm_1$ production and
decay at $\roots = 182.7$~GeV are shown by
dashed lines.
The light shaded areas  
show the additional exclusion regions due to
direct neutralino searches and other OPAL search results (see text).

}
\label{masslimc}
\end{minipage}
\end{figure}

\begin{figure} \centering
\begin{minipage}{16.0cm}
\begin{center}\mbox{
\epsfig{file=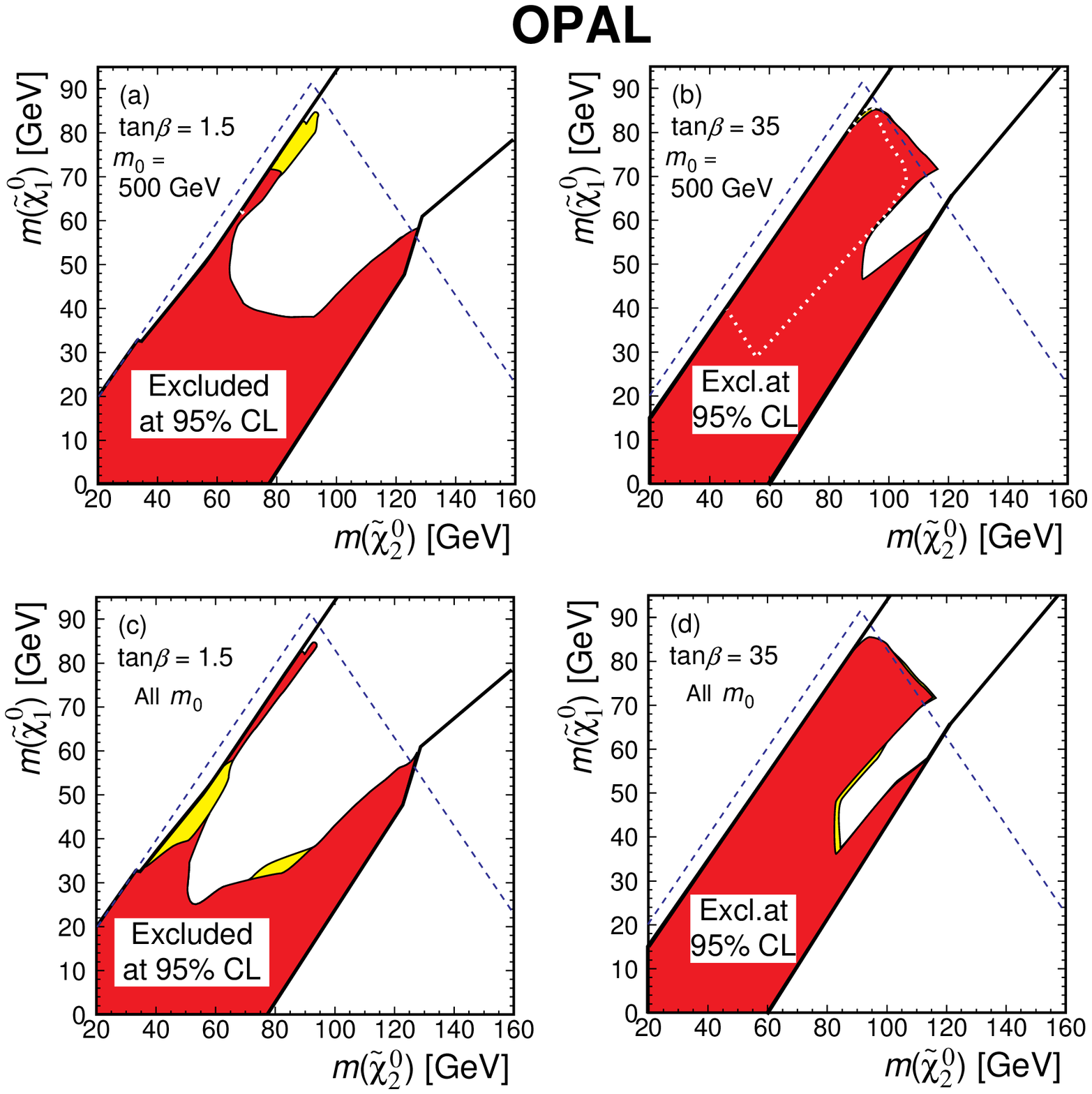,width=16.0cm}
}\end{center}
\vspace{-9mm}
\caption[]
{
The 95\% C.L. excluded region
in the ($m_{\nt_1}$,$m_{\nt_2}$) plane
within the framework of the CMSSM with $m_0 \geq 500$~GeV
for (a) $\tan \beta=1.5$ and for (b) $\tan \beta=35$.
The excluded region valid for all $m_0$  
for (c) $\tan \beta=1.5$ and for (d) $\tan \beta=35$.
The thick solid lines represent the theoretical bounds
of the CMSSM parameter space as defined in the text.
The kinematical boundaries for $\nt_2 \nt_1$ production and
decay at $\roots = 183$~GeV are shown by
dashed lines.
The regions excluded outside of the kinematical boundary
$m_{\nt_2} + m_{\nt_1} = \roots$ is due to 
the interpretation of the $\chp_1 \chm_1$ search results.
The dark region is excluded by the results of direct $\chp_1 \chm_1$ searches.
The light shaded areas  
show the additional exclusion regions due to
direct neutralino searches and other OPAL search results (see text).
The regions inside the white dotted lines in Fig.(a) and (b) 
would be excluded by the direct neutralino searches alone.
}
\label{masslimn}
\end{minipage}
\end{figure}

\begin{figure} \centering
\begin{minipage}{16.0cm}
\begin{center}\mbox{
\epsfig{file=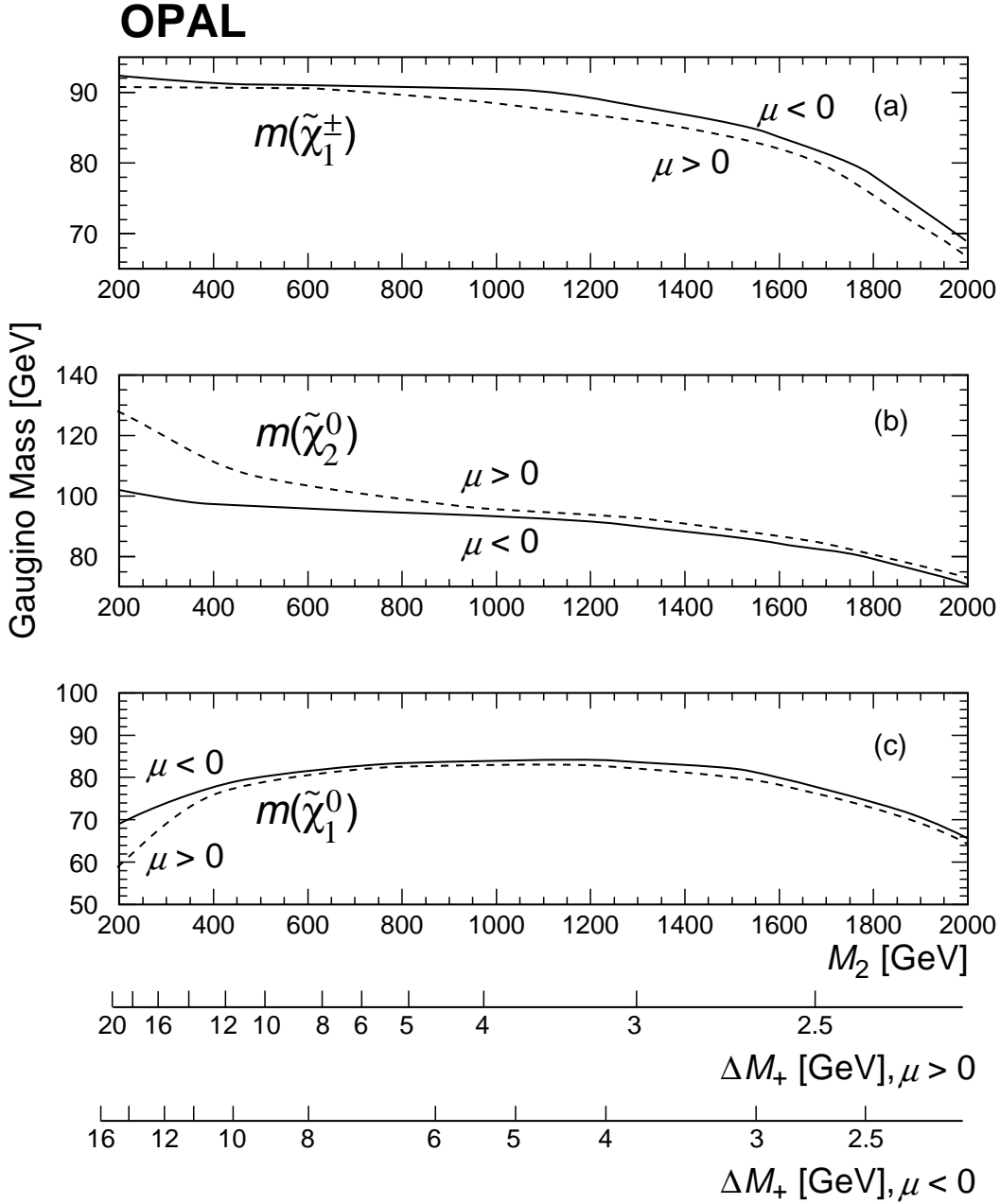,width=14.0cm}
}\end{center}
\vspace{-9mm}
\caption[]
{
The 95\% C.L. lower mass limits for
(a) $\chp_1$, (b) $\nt_2$, and (c) $\nt_1$ for
$\tan\beta = 1.5$ and $m_0 = 500$~GeV for slices of constant
value of $M_2$.  Limits are shown separately for 
$\mu < 0$ (solid lines) and $\mu > 0$ (dashed lines).
Curves for larger values of $\tan\beta$ are in general 
between those shown for $\mu < 0$ and $\mu > 0$.
The corresponding values of mass difference $\Delta M_+$ between the 
chargino and the lightest neutralino are also shown on scales
below the plots.
}
\label{plotm2}
\end{minipage}
\end{figure}

\begin{figure} \centering
\begin{minipage}{16.0cm}
\begin{center}\mbox{
\epsfig{file=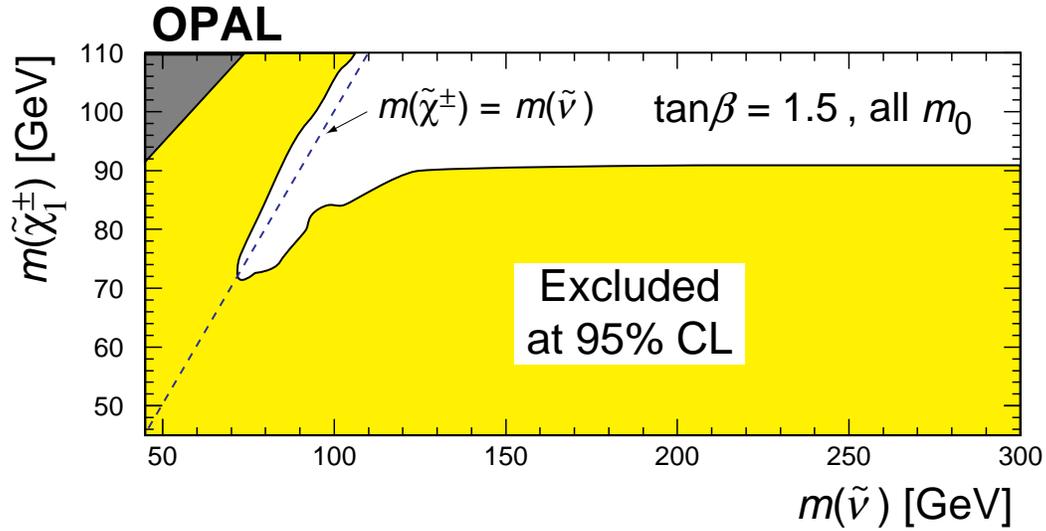,width=14.0cm}
}\end{center}
\vspace{-9mm}
\caption[]
{
The 95\% C.L. excluded region in the ($m_{\snu}$, $m_{\chp_1}$) plane
for $\tan\beta = 1.5$ and for all $m_0$ values.
The region close to the line of $m_{\snu} = m_{\chp}$
where the $\chp_1 \chm_1$ search fails due to small visible energy 
is excluded using the limits for sneutrinos and OPAL
limits for sleptons. The dark shaded region is theoretically inaccessible.
}
\label{limchsnu}
\end{minipage}
\end{figure}

\begin{figure} \centering
\begin{minipage}{16.0cm}
\begin{center}\mbox{
\epsfig{file=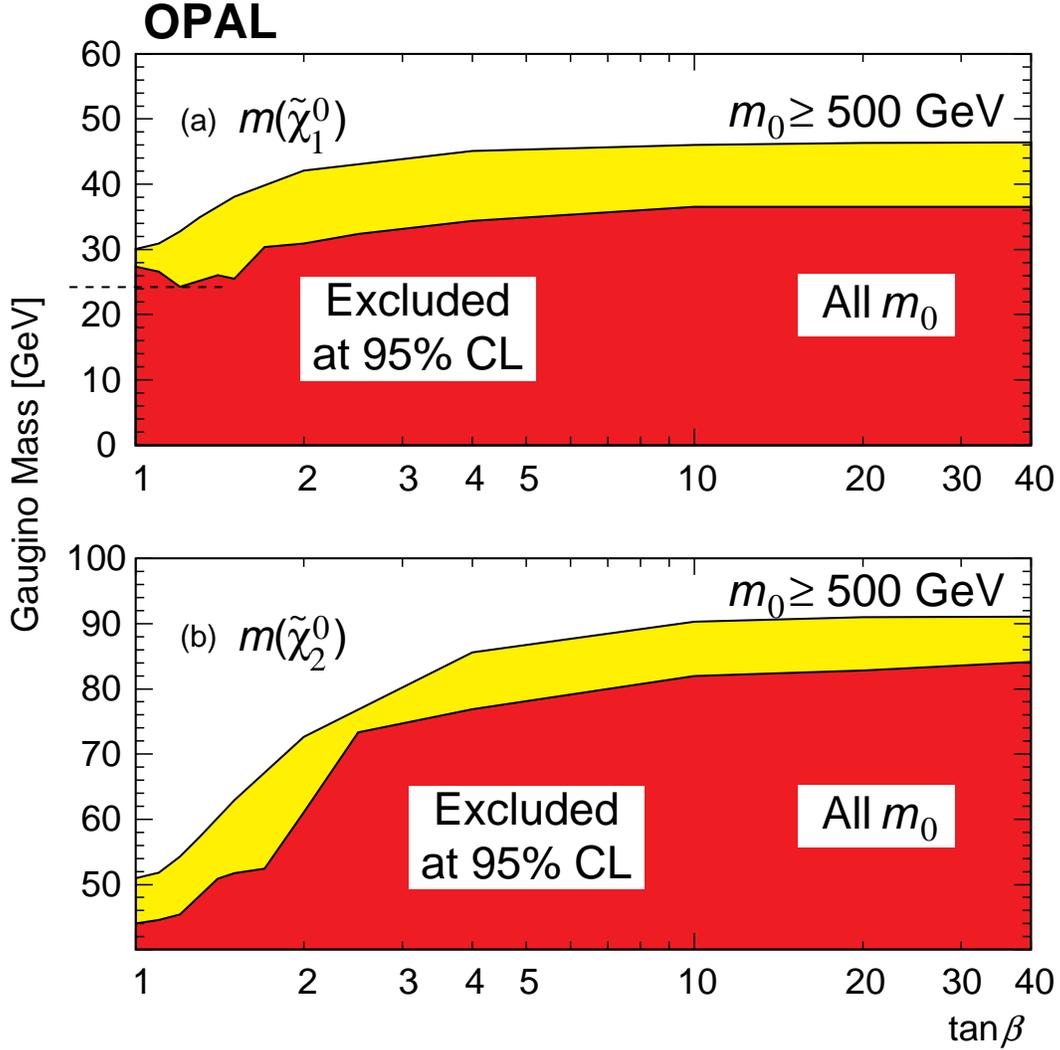,width=14.0cm}
}\end{center}
\vspace{-9mm}
\caption[]
{
The 95\% C.L.  mass limit on
(a) the lightest neutralino $\nt_1$ and
(b) the second-lightest neutralino $\nt_2$ as a function of 
$\tan\beta$ for $m_0 \ge 500$~GeV.  The mass limit on
$\nt_2$ is for the additional requirement of $\Delta M_0 > 10$~GeV.
The exclusion region for $m_0 \ge 500$~GeV is shown
by the light shaded area and the excluded region valid
for all $m_0$ values by the dark shaded area.
}
\label{limtanb}
\end{minipage}
\end{figure}
\end{document}